\documentclass[12pt]{article}
\usepackage{newinutile}
\usepackage{amsmath,amssymb}
\usepackage{epsfig,graphics,color,calc,graphicx}
\usepackage{cite}
\usepackage{mathrsfs}
\usepackage[title]{appendix}

\newcounter{enmccommento}

\newcommand{\newatop}[2]{\genfrac{}{}{0pt}{}{#1}{#2}}

\begin{document}
\title{Deterministic reversible model of non-equilibrium phase transitions 
and stochastic counterpart}

\author{Emilio N.\ M.\ Cirillo}
\email{emilio.cirillo@uniroma1.it}
\affiliation{Dipartimento di Scienze di Base e Applicate per l'Ingegneria,
             Sapienza Universit\`a di Roma, 
             via A.\ Scarpa 16, I--00161, Roma, Italy.}

\author{Matteo Colangeli}
\email{matteo.colangeli1@univaq.it}
\affiliation{Dipartimento di Ingegneria e Scienze dell'Informazione e 
Matematica,
Universit\`a degli Studi dell'Aquila, via Vetoio, 67100 L'Aquila, Italy.}

\author{Adrian Muntean}
\email{adrian.muntean@kau.se}
\affiliation{Department of Mathematics and Computer Science,
Karlstad University, Sweden.}

\author{Omar Richardson}
\email{omar.richardson@kau.se}
\affiliation{Department of Mathematics and Computer Science,
Karlstad University, Sweden.}

\author{Lamberto Rondoni}
\email{lamberto.rondoni@polito.it}
\affiliation{Dipartimento di Scienze Matematiche, 
Politecnico di Torino, corso Duca degli Abruzzi 24, I--10129 Torino, Italy.}
\affiliation{INFN, Sezione di Torino, Via P. Giuria 1, 10125 Torino, Italy.}

\begin{abstract}
$N$ point particles move within a billiard table made of two circular cavities connected by a 
straight channel. The usual billiard dynamics is modified so that it 
remains deterministic, phase space volumes preserving and time reversal invariant. Particles 
move in straight lines and are elastically reflected at the boundary of the table, as usual, 
but those in a channel that are moving away from a cavity invert
their motion (rebound), if their number exceeds a given threshold $T$.
When the geometrical parameters of the billiard table are fixed, this mechanism gives rise to non--equilibrium phase transitions in the large $N$ limit: 
letting $T/N$ decrease, the homogeneous particle distribution abruptly turns 
into a stationary inhomogeneous one. 
The equivalence with a modified Ehrenfest two urn model, motivated by the ergodicity of 
the billiard with no rebound, allows us to obtain analytical results that accurately 
describe the numerical billiard simulation results. Thus, a stochastic exactly solvable 
model that exhibits non-equilibrium phase transitions is also introduced.
\end{abstract}



    

\maketitle

\section{Introduction}
\label{s:introduzione}
\par\noindent
Phase transitions among equilibrium states are widely investigated and 
well understood phenomena, which is one of the
major achievements of statistical mechanics \cite{PaLeMe,ChRoVu}. Non-equilibrium phase transition 
\cite{Lebo85,Henkel,Gaveau,Evans00,Lubeck,EvansHanney05,Igloi}, 
on the other hand, are much less investigated and currently, a comprehensive framework seems to be lacking, as it is
in general the case for non-equilibrium statistical mechanics \cite{MeRoNon}. However, the past 
three decades have witnessed important advances, based on the generalization of equilibrium 
fluctuations theory and its consequences, such as the fluctuation-dissipation theorem and response theory
\cite{EvMo,BePuRoVu,CoRoVe,Sekimoto}. While stochastic models have occupied the largest fraction of the
recent specialized literature, because simpler to treat and more inclined to produce results \cite{Kurchan}, 
major achievements came from the study of deterministic time reversal invariant (TRI) dynamical and dissipative,
{\em i.e.}\ phase space volumes contracting, systems, 
such as those of non-equilibrium molecular dynamics (NEMD). 
These advances 
include relations between dynamical quantities such as the Lyapunov exponents and macroscopic
properties such as the transport coefficients \cite{ECM1}, fluctuation relations
\cite{ECM2}, linear response relations for perturbations of non-equilibrium steady states
\cite{Ruelle,CoRoVu,DemRo}, and exact response relations together with novel ergodic 
notions \cite{EvSeWi,Typicality,JeRo2016}. 
It is indeed more natural to investigate in deterministic reversible dynamics, rather than in stochastic 
processes, the properties related to microscopic reversibility. In fact, stochastic processes are 
intrinsically irreversible, although they may enjoy the property known as detailed 
balance \cite{Kreuzer}.

Concerning non-equilibrium phase transition, most results are obtained for stochastic processes.
In fact, various kinds of abrupt transitions have been reported also
in the NEMD literature. They have been observed in systems of 
small numbers of particles, when dissipation is increased; see, {\em e.g.}\ ``string phases'' in 
shearing fluids, where dissipation can be so strong that 
chaos of fluid particles is damped and 
ordered phases arise \cite{EvMo,Erpenbeck}. The non-equilibrium Lorentz and Ehrenfest gases are 
even more striking from this point of view, because 
chaos in the dynamics of non--interacting particles can be 
tamed by dissipation, and an impressive variety of 
bifurcation--like and hysteresis--like phenomena 
may result, cf.\ \cite{PRE94,CHAOS95,PRE95,CHAOS09}. However, these behaviors have not been investigated 
as transitions that occur in some kind of macroscopic limit, or for conservative dynamics.
Therefore, the question arises whether they can be obtained in deterministic, TRI and possibly
non--dissipative systems. 

\begin{figure}[h]
\centering
\includegraphics[width = 0.6\textwidth]{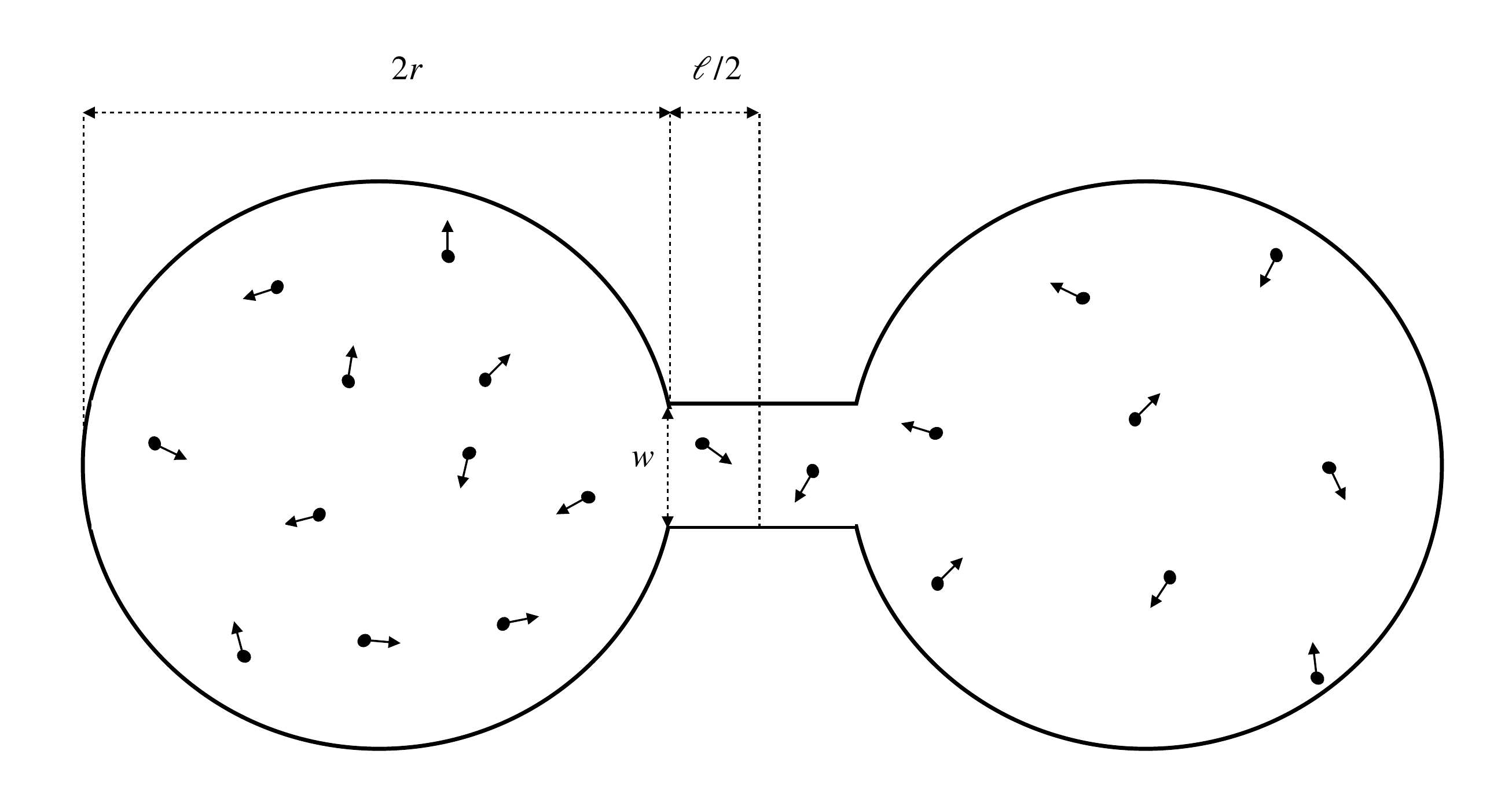} 
\caption{Schematic representation of the model.}
\label{fig:model}
\end{figure}

In the present paper, we investigate the onset 
of non-equilibrium phase transitions in a conservative, TRI
dynamical system of phase space $\cal M$, consisting of $N$ point particles moving in straight 
lines at constant speed $v=1$, within a billiard table $\Lambda$ made of two circular \emph{urns} 
of radius $r$, connected by a rectangular channel of width $w$ and length $\ell$, cf.\ Fig.\ref{fig:model}. 
The channel is then divided in two parts: its left half 
$G_\textup{L}$ and its right half
$G_\textup{R}$, each of length $\ell/2$,
called \emph{gates}. When particles hit the
boundary $\partial \Lambda$ of the table, they are elastically reflected. This means that their speed is preserved
while their velocity is reversed so that the outgoing angle with respect to the normal to $\partial\Lambda$ at the
collision point equals the incoming angle. 
So far, we have described a standard ergodic billiard \cite{Bunimovich,Bunim05}. 
We then add one further dynamical rule:
when the total number of particles in any of the two gates that point towards the other gate exceeds the threshold
value $T$, the horizontal component of the velocity of those particles 
is reversed. When such a \emph{bounce--back} mechanism takes place, the 
particles in the interested gate directed towards the other gate go back to the 
urn from which they came. The particles that are coming from the other urn continue unaltered 
their motion. 

As initial condition, a certain fraction of particles is placed in the left urn and the remaining fraction in the 
right one; their positions and directions of velocities are chosen at random with uniform probability. 

If $T \ge N$, one has the usual TRI and ergodic billiard dynamics.
Ergodicity implies that each particle spends an equal amount of time in the two urns, hence, for large $N$, 
an equilibrium distribution of particles is reached, in the sense that approximately the same number of 
particles are found in the urns, apart from small oscillations about that number, and apart from very rare 
large deviations related to correspondingly long recurrence times. 
Time reversal invariance means that there exists an involution $i : {\cal M} \to {\cal M}$ of the \emph{phase space} 
in itself, that anticommutes with the time evolution $S^t : {\cal M} \to {\cal M}$, so that \cite{JeRo2016}:
\begin{equation}
S^t i \Gamma = i S^{-t} \Gamma \,  \quad  \text{for all} ~ \Gamma \in \cal M
\label{TRI}
\end{equation}
where $t \in \mathbb{R}$ is the time, and $i^2$ is the identity operator on $\cal M$. For instance, denoting
by {\bf q} the positions of the particles, and by {\bf p} their momenta, so that $\Gamma=({\bf q},{\bf p})$  
is a phase in $\cal M$, the usual reversal operation is defined by 
$i ({\bf q},{\bf p}) = ({\bf q},-{\bf p})$. However, various other involutions could be considered, see e.g. \cite{Bonella,Coretti}.

If $T < N$, ergodicity implies that sooner or later a number of particles larger than $T$ will be found in, say,
$G_\textup{L}$ with velocity pointing towards $G_\textup{R}$. 
At that time, the standard billiard dynamics will
be interrupted, and the particles in the left gate that were going rightward will be reflected
as if they had hit a rigid vertical wall: this event does not alter the reversibility of the 
dynamics, as the usual involution that preserves positions and inverts momenta works also in this case.
Indeed, the bounce--back mechanism within a gate is like an elastic collision with a wall: the only 
difference is that we cannot trivially see such a wall, because it occupies a precise region $W \subset \cal M$ 
of the phase space, but not of the real two dimensional space. Nevertheless, $W$ exists and is 
identified by the condition that more particles than $T$ lie in a gate with velocity pointing 
towards the other gate. 

The region $W$ can be considered as removed from the phase space, like the region corresponding to particles inside 
a scatterer of the phase space of the Lorentz gas. Analogously, the boundary of $W$ 
acts like the one corresponding to the surface of the Lorentz gas scatterer: particles equally preserve their energy, 
and their velocities are equally elastically reflected.
Indeed, suppose the $G_\textup{L}$ contains $T$ particles moving towards 
$G_\textup{R}$, while one 
more particle is entering. As the number of such particles inside 
$G_\textup{L}$ turns $T+1$, 
their motion is inverted, and they move back to the left urn. The last particle entering $G_\textup{L}$
spends only a vanishing time inside that gate, while the other particles remain in $G_\textup{L}$ for as 
long as they had been, since their horizontal speed is the same before and after bouncing back. 
Our main result about this conservative TRI particle system, when the geometrical parameters of the billiard table are fixed, is that: 

{\em a non-equilibrium phase transition takes place, for a given $T/N$, in the large $N$ limit}. 

\noindent
The transition consists in switching from a state in which half of the particles lies in the left half 
and the rest in the right half of $\Lambda$, to the state in which almost all particles lie in one of 
the two halves. The non-equilibrium nature of the inhomogeneous state is 
revealed by the fact that it rapidly relaxes to the homogeneous equilibrium state if the rebound 
mechanism is switched off. 

The reason for the phase transition is that large $N$ and small $T/N$ 
make it harder for particles in the 
urn with higher density to reach the urn with lower density, since the 
bounce--back mechanism is more 
frequent at higher density. Therefore, particles exiting the urn with low density reach the urn with high 
density and remain trapped, while the difference between the densities of the left and right urns
grows. Differently, when $N$ is large but $T/N$ is not small, the density in the two urns tends 
to equalize and to establish a homogeneous state.\footnote{Possibly neglecting the state in the gates.} 
For sufficiently large $N$, 
this scenario is confirmed by our simulations of the billiard dynamics.

This phenomenon must be properly interpreted. 
For any finite $N$, finite recurrence times make the system explore in time many
different distributions of particles, apart from those forbidden by the 
bounce--back rule. 
As $N$ grows, such fluctuations become
so rare compared to any physically relevant time scale, that a given 
state can be legitimately taken as stationary. 
This phenomenon is analogous to that concerning the
validity of the $H$--theorem of the kinetic theory of gases. While no real gas is made of infinitely many molecules,
hence the corresponding $H$--functional in principle is not monotonic and is affected by recurrence, the 
Boltzmann equation and the $H$--theorem perfectly describes such systems \cite{Cercig,Castigl}. When $N$ grows, fluctuations in the monotonic behavior of $H$ become relatively smaller and the recurrence times longer, so that for a macroscopic system, deviations from the behavior predicted by the Boltzmann 
equation are not expected within any physically relevant time scale.

Thus, unlike rarefied gases that have a single stationary state -- the equilibrium state --
 our system may be found in a 
``polarized'' steady state, in which most of the particles are gathered in one urn, or in a ``spread'' (homogeneous) steady state, with
same numbers of particles in each urn. When the geometrical parameters of the billiard table are fixed, the 
onset of either of 
the two states depends on $N$ and on $T/N$, and 
stationarity must be intended as in the kinetic theory of gases.

The fact that $N$ cannot exceed $T$ inside a gate produces a state
with a number density that is lower in the gates than in the urns. 
A gate may at most host 
2$T$ particles, $T$ going towards 
and $T$ coming from the other gate. 
That corresponds to a number 
density $4T/\ell w$, which can be arbitrarily smaller than the highest possible density in one 
urn,  $N/\pi R^2$, and than the density in the urns when they are equally populated, which approximately
equals $N / 2 \pi R^2$. In other words, given the 
table $\Lambda$ and the threshold $T \ge 1$, a sufficiently 
large $N$ makes the equilibrium state impossible: 
both the aforementioned states, i.e., the homogeneous and the 
polarized one, are going to be non-equilibrium 
steady states.\footnote{One might think that this is analogous to the case of 
Knudsen gases \cite{DGM},
obtained when two containers are connected by a 
capillary thinner than the particles mean free path. 
As particles
do not interact within that channel, 
they do not bring information about the thermodynamic state from one container 
to the other. Consequently, the system does not relax to a state 
of equal temperature and pressure but,
denoting by $T_i, P_i$, $i=\textup{L},\textup{R}$ 
the temperature and pressure in the left and the right container, the states
obeying
\begin{equation}
{P_\textup{L} \over \sqrt{T_\textup{L}}} 
= 
{P_\textup{R} \over \sqrt{T_\textup{R}}}
\label{Knudsen}
\end{equation}
are all stationary. Despite some analogy, this is not our case. In fact, our particles do not interact at all, 
hence thermodynamics does not apply to them.
}

This sheds light on the relation between microscopic reversibility and phase space
volume preserving property of our dynamics, and the realization of non-equilibrium steady 
states and phase transitions. Indeed, our case seems to be different from those reported 
in the existing literature. 
In the first place, note that the effects of microscopic reversibility may be verified in certain phenomena, 
even if the standard time reversal symmetry does not hold, because alternative {\em equivalent} symmetries 
do, cf.\ {\em e.g.}\ \cite{Bonella,Coretti}. Such an equivalence depends on the observables of interest and
on the relevance of statistics. Therefore, in certain situations reversibility may even be totally absent,
without affecting properties generally associated with reversibility, such as the validity of the fluctuation 
relations \cite{CKDR,ColRon}. In NEMD, TRI microscopic dynamics 
is associated with an irreversible contraction
of phase space volumes, that in some cases may be quite drastic, and be accompanied by abrupt collapse of 
the phase volumes dimensions 
\cite{Erpenbeck,PRE94,CHAOS95,PRE95,CHAOS09,BGaxiom,MRaxiom}.

Our dynamics, on the other hand, is TRI according to the standard reversal operation, and it is also 
phase space volumes preserving, although it prevents equilibrium. It seems that our case is similar to 
the one described in Ref.\cite{Plathe}, concerning a molecular dynamics algorithm for simulations of a shearing 
fluid. In that case, a kind of Maxwell demon exchanges some fast and slow particles. When a Maxwell 
demon acts, some thermodynamic rules appear to be violated but, in reality, rather energetic environments 
must operate, greatly dissipating, in order to produce such a ``violation'' consistently with
thermodynamics \cite{Masuyama}. Our bounce--back mechanism, which defeats the trend towards equilibrium 
of the standard billiard \cite{Bunimovich}, is an analogous mechanism.

As in other cases, the statistical nature of the quantities of our interest justifies the introduction 
of a stochastic counterpart of our deterministic model, which allows a detailed mathematical detailed analysis 
not easily accessible in the deterministic framework. In general, associating stochastic processes to deterministic 
systems has proven quite useful; for instance, it has been the key, 
via representations of SRB measures, 
to results such as the fluctuation relation \cite{ECM2,GC}. Therefore, the second part of this paper is 
devoted to a stochastic two urn model that is inspired by the deterministic model, and for which the 
existence of a non-equilibrium phase transition can be investigated analytically. We remark that thresholds 
affecting particles dynamics have proven effective in other investigations of stochastic models such as 
those of Refs.\cite{CCCMpsy,CCMbottleneck,CCMpre2016}. We also observe that phase transitions are found
in stochastic systems similar to ours, such as the Ehrenfest urn model with interactions \cite{Tseng,Cheng}.
In fact, our stochastic model 
reduces to 
the classical Ehrenfest urn model, when the bounce--back mechanism
is inhibited, and the system asymptotically relaxes to equilibrium \cite{OrbBell67}.

To match the deterministic and the stochastic models, quantities such as the frequency of the bounce--back 
events are required. For billiard systems, these quantities can be estimated considering the ergodicity of the 
standard billiard, and the fact that it yields more accurate results for larger $N$. Initially,  the dynamics appears like 
that of an ergodic billiard with a hole \cite{Bunim05,BunYur}. Then, if the number of
particles in one urn is large, the adjacent gate is rapidly filled with particles moving towards the other
urn, which then bounce back. Therefore, the larger $N$, the less likely for particles to leak out of one urn.
At the same time, motion inside the urns is chaotic, which tends to produce uniform space and velocity 
distributions, justifying a probabilistic approach. 
In Section \ref{s:modres}, these calculations are carried out, and are shown to 
accurately describe the dynamics.

\section{The deterministic model}
\label{s:modres}
\par\noindent
This section is devoted to the study of the deterministic billiard model. 
A preliminary heuristic discussion will be followed by the 
numerical study and by some analytical interpretations of the results.

\subsection{Polarized and homogeneous states. Outlet and leaking currents}
\label{s:hgue}
\par\noindent
The motion of particles inside each reservoir is ergodic, 
therefore, for large $N$, we expect that particles leave an urn to enter the adjacent gate with a rate 
that simply depends on $N$, on 
the radius of the urn, on the width of the gate and on the speed of the particles. 
If the rate is low, a given threshold $T$ may 
not be exceeded by the number of particles in the gate,
and particles safely cross the channel towards the opposite urn. 
On the other hand, if such a rate is high with respect to the typical 
time needed by particles to walk through the gates, $T$ can be frequently exceeded,
making particles bounce back. 

Starting from a configuration in which the two reservoirs share the same (large) number of particles, 
we have two extreme situations to consider for stationary states:

\noindent
{\bf i)} If the ratio $T/N$ is large,
the bounce--back mechanism is not
effective. Particles move freely from one reservoir to the other 
and, at stationarity, the number of particles in the two urns is
approximately constant and equal. This state is stable and will 
be referred to as a \emph{homogeneous} state. 
In this case, relatively large average currents flowing in opposite directions
from one urn to the other, that we call \emph{outlet currents}, balance each other.

\noindent
{\bf ii)} If the ratio $T/N$ is small, the particles 
inside a gate may frequently exceed $T$ and bounce back 
to their original urn. The corresponding average currents, which we call \emph{leak currents},
are small, hence the initial condition with an equal number of particles in the two urns
is only slightly perturbed by them. However, even a small fluctuation in this distribution 
of particles, due to an instantaneous unbalanced leak current, will lead to a different 
bounce--back frequency in the two urns, which will in turn amplify the difference in the number 
of particles in the two urns, till an inhomogeneous (polarized) state will be achieved.

To distinguish between the different stationary states, 
we introduce an order parameter $\chi$, called \emph{mass displacement}, that is the absolute value of 
the difference between the time averaged number 
of particles in the two halves of the table $\Lambda$, 
divided by $N$.
The parameter $\chi$ is thus close to one in the polarized state and 
close to zero in the homogeneous state. 
Note that in the appropriate region of the parameter space, even when starting from the homogeneous state (actually, the initial datum is irrelevant), the system ends up in the polarized state. This can be seen as an instance of a \textit{transient} uphill mass transport phenomenon, in which particles are observed to preferably move from regions of lower concentration to regions of higher concentration \cite{CdMP1,CdMP2,CdMP3,CC17}.

\subsection{Efficiency of the bounce--back mechanism}
\label{s:efreg}
\par\noindent
Here we estimate 
the values of the parameters for which the bounce--back mechanism 
becomes effective. Thus, for a fixed total number of particles $N$,
we derive Equation \eqref{det010}, which identifies, in the parameter space
$r$--$w$--$\ell$--$T$, the hypersurface separating the region in which 
the threshold mechanism is efficient from the region in which it is not.

The idea is the following: relying on the ergodicity argument described above,
and assuming very large $N$, we consider the homogeneous state and compare the 
rate at which particles enter a gate as well as the typical 
time needed to cross the gate. Take $\delta$ such that $v\delta\ll w$, recalling that $v=1$. 
The probability that the particle enters a gate from the adjacent urn in a time smaller than 
$\delta$ is $p_\delta = 2w v^2 \delta/(2\pi v\cdot\pi r^2)=w v \delta/(\pi A)$
\cite[Appendix~A]{CCCFP2016}, where 
\begin{equation}
\label{det0-10}
A = A(r,w)=
\pi r^2-r^2\arcsin\Big(\frac{w}{2r}\Big)+\frac{1}{4}w\sqrt{(2r)^2-w^2}
\end{equation}
is the area of the urn. Obviously, $A \approx \pi r^2$ when $w$ is 
small compared to the radius $r$ of the urn.
Requiring $p_\delta\sim1$, we get $\pi A/(w v)$ as an estimate of the
typical time to exit the reservoir.

The average 
horizontal component of the particles entering the channel is 
$2v/\pi$. Hence, the typical time to cross the gate 
is $(\ell/2)/(2v/\pi)=\ell\pi/(4v)$.

Consider the time $t$ and the small interval $\delta$.  
Each of the $N/2$ particles in the reservoir tries to enter the channel 
in the small time $\delta$. It is like performing $(N/2)(t/\delta)$ 
Bernoulli trials with success probability 
$w v \delta/(\pi A)$. 
Hence, the number of particles entering the channel from one 
of the two reservoir during the time $t$ is estimated by 
$t(N/2)w v/(\pi A)$.
More in general, we define, for later use, the \emph{outlet 
current} 
$J_\textup{o}(n)=nwv/(\pi A)$
coming from an urn with $n$ particles. 


Since on average particles take about the time $\ell\pi/(4v)$
to cross the gate, the typical number of particles 
inside a gate is equal to the number of particles 
which enter it in the time $\ell\pi/(4v)$, namely, 
$\ell\pi/(4v)(N/2)w v/(\pi A)=Nw \ell/(8A)$.

We conclude that the bounce--back mechanism is efficient 
provided the parameters of the model satisfy the inequality
\begin{equation}
\label{det000}
\frac{Nw\ell}{8A}>T
\;\;.
\end{equation}
Thus, exploiting \eqref{det0-10}, we expect that the region
of the parameter space $r$--$w$--$\ell$--$T$ 
in which the threshold mechanism is efficient and that in which it is not 
are separated by the hypersurface 
\begin{equation}
\label{det010}
Nw\ell=T 8
\Big[
     \pi r^2-r^2\arcsin\Big(\frac{w}{2r}\Big)+\frac{1}{4}w\sqrt{(2r)^2-w^2}
\Big]
\,.
\end{equation}

\subsection{Numerical results}
\label{s:num-din}
\par\noindent
To support numerically this intuition, we simulated the model and measured the stationary average 
value of the mass displacement
\begin{equation}
\chi = {\left| N_R - N_L \right| \over N},
\label{chi}
\end{equation}
where $N_L$ is the number of particles in the left half of $\Lambda$, and
$N_R$ that of the particles in its right half, with $N_L+N_R=N$. In particular, a 
random initial condition $\Gamma \in \cal M$ is used, and the standard 
event--driven 
algorithm is implemented, recording the phase whenever a particle collides with the 
boundary $\partial \Lambda$, reaches the entrance of a gate or crosses from one gate 
to the other. This is necessary in our case, in order to verify whether the particles 
in a gate going towards the other gate exceed $T$, or do not. 

A simulation is first performed up to a number $M_{\rm t}$ of events that are sufficient 
for the instantaneous value of $\chi$ to become constant
(modulo fluctuations), {\em i.e.}\ until a stationary 
state appears to be reached.
This requires large $N$, and we have found that $N=O(10^3)$ suffices. 
The simulation stops when a total number $M_\text{f}$ of events
is reached. Then, $\chi$ is averaged over the time corresponding to the final 
$M_\text{f} - M_{\rm t}$ steps; an ensemble of such averages is obtained simulating the
evolution from different initial conditions; and finally the ensemble average of the 
time averages is produced.\footnote{The ensemble average is only performed for numerical 
efficiency, since we have observed that a single very long simulation would suffice.}

\begin{figure}[t]
\centering
\includegraphics[width = 0.99\textwidth]{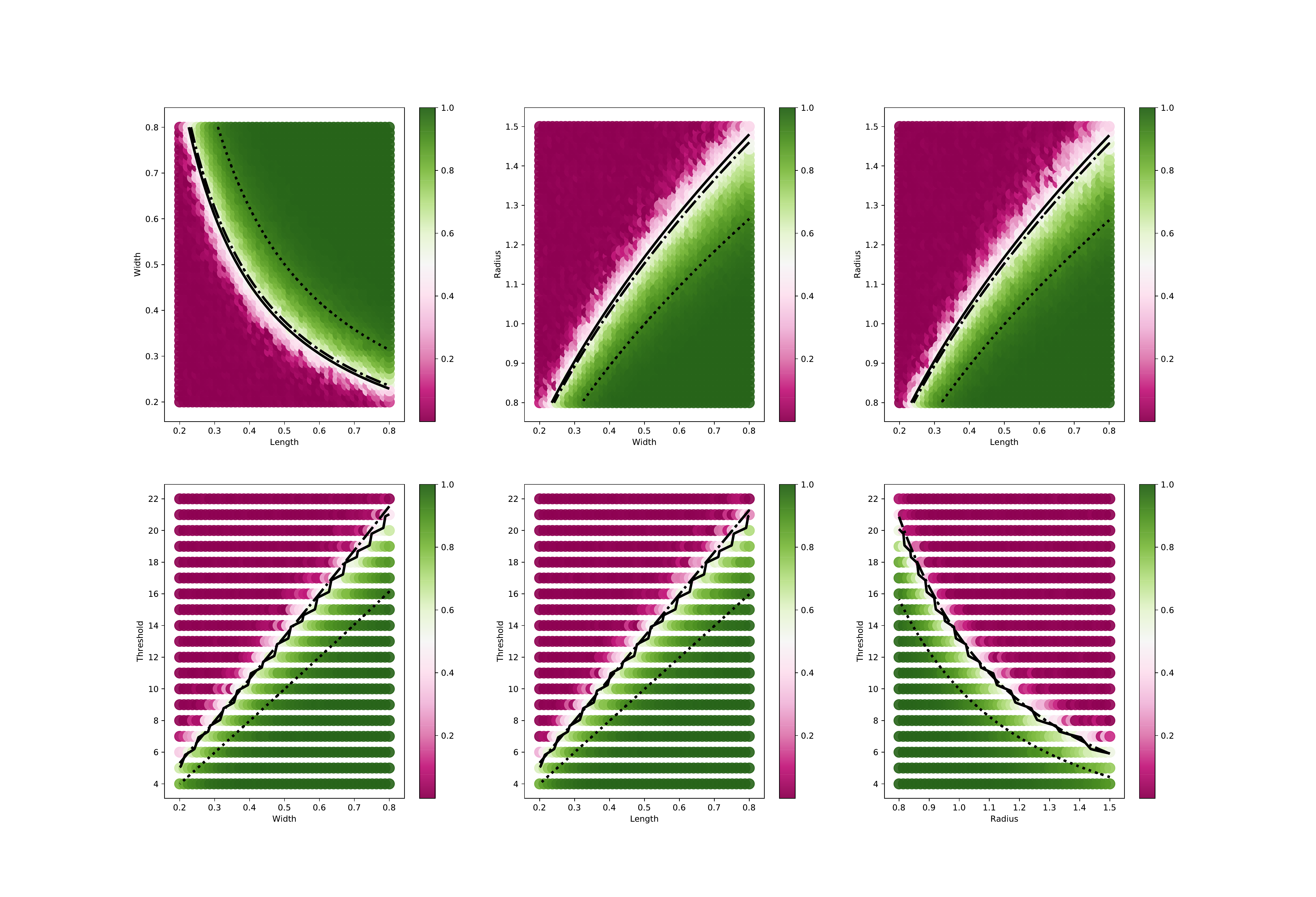} 
\caption{Average mass displacement $\chi$ for $N=10^3$, as a function of two 
parameters, with the remaining ones fixed to a reference value. 
The reference values are 
$r=1$, $T=10$, $w=0.5$, and $l=0.5$.
The black dotted line is the theoretical curve \eqref{det010}.
The black dashed dotted line is obtained by multiplying 
the right hand side of the theoretical curve \eqref{det010} by 
the fitting parameter $0.75$.
The black solid line is the curve \eqref{eq:tl060}.
}
\label{fig:ehr00}
\end{figure}

For sake of comparison, our simulations have been collected in classes in which two parameters are fixed 
to some reference value, while the remaining pair of parameters varies in a relatively wide range of values.
In Figure~\ref{fig:ehr00}, we have thus plotted our results for $N=10^3$.
For each value of the parameters we have considered one single realization of the dynamics, namely, the 
evolution of the system starting from one random initial configuration $\Gamma$ with mass spread 
$\chi=0.50$. 
The reference value for the four parameters are the following $r=1$, $T=10$, $w=0.5$, and $\ell=0.5$.  
The simulation has been performed with $M_\text{t}=5\times10^6$ and 
$M_\text{f}=6\times10^6$. We have 
performed the same simulations starting from an initial configuration 
with mass spread equal to $0, 0.25, 0.75, 1$ and we found perfectly 
similar results. 

The parameter region has been chosen around the 
curve defined by the equation \eqref{det010} and derived in 
Subsection~\ref{s:efreg}.
Indeed, if all the 
parameters but two are fixed, \eqref{det010} is the algebraic 
equation of a curve in the plane in terms of the two varying parameters. 
We notice that such a curve (black dotted line in the 
Figures~\ref{fig:ehr00}) discriminates the region in which the 
bounce--back mechanism is efficient and that in which it is not, thus
locating a rather sharp transition from polarized to homogeneous states. 
In our figures, the green region corresponds to a polarized state ($\chi=1$),
while the deep purple region corresponds to a homogeneous state ($\chi=0$).

In all the plots, the black dotted 
theoretical curve lies closer to the green part of the graph,
which is the inhomogeneous steady state.
Hence, though the level curves of $\chi$ rather closely follow 
the theoretical curve,  
the theoretical prediction underestimates the bounce--back mechanism corresponding to 
finite $N$ and $w$.
Indeed, according to the inequality \eqref{det000}, 
in the plane $\ell$--$w$ the region in which the bounce--back mechanism is efficient lies 
above the curve, whereas in our pictures, except one, it lies below it. 
As noted above, this is not a surprise, since \eqref{det000} is not 
a sharp mathematical inequality, but rather an informed guess, whose accuracy should 
increase with $N$, as long as ergodicity can be legitimately invoked. 
As a matter of fact, we observe that the numerically computed interface 
differs from the theoretical one just by a multiplicative factor, 
cf.\ the dashed dotted black curve in Figure~\ref{fig:ehr00}, which is the theoretical 
curve multiplied by a number.

\begin{figure}[t]
\centering
\includegraphics[width = 0.99\textwidth]{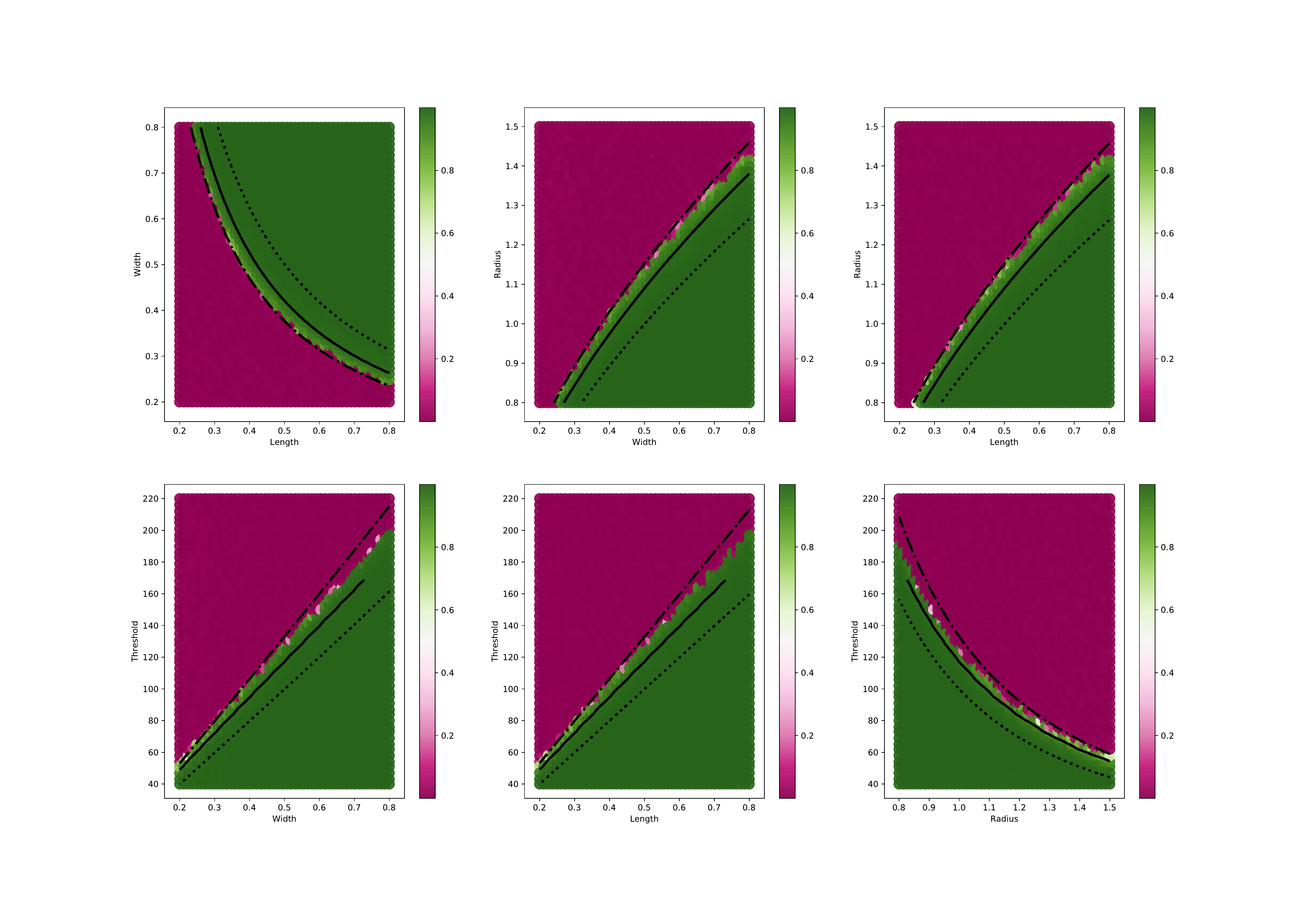} 
\caption{Mass displacement $\chi$ as a function of two parameters with the 
remaining ones fixed to their reference value. The reference values are 
$r=1$, $T=100$, $w=0.5$, and $\ell=0.5$.
The total number of particles is $N=10^4$.
The black dotted line is the theoretical curve \eqref{det010}.
The black dashed dotted line is obtained by multiplying 
the right hand side of the theoretical curve \eqref{det010} by 
the fitting parameter $0.75$.
The black solid line is the curve \eqref{eq:tl060}.
}
\label{fig:ehr02}
\end{figure}

In Figure~\ref{fig:ehr02}, we have plotted our results for $N=10^4$.
For each value of the parameters we have considered 
one single realization of the dynamics
starting from a randomly chosen 
initial configuration
with initial mass spread 
$0.50$. 
The reference value for the four parameters are 
$r=1$, $T=100$, $w=0.5$, and $\ell=0.5$.  
The simulation has been performed with 
$M_\text{t}=5\times10^6$ 
and 
$M_\text{f}=6\times10^6$.
Results are similar to those obtained in the case with smaller values 
of $N$. As before, we have 
performed the same simulations starting from an initial configuration 
with values of mass spread equal to $0, 0.25, 0.75, 1$ and we again found perfectly 
similar results. 

\subsection{Computing the transition line}
\label{s:tranl}
\par\noindent
We now develop a theoretical argument to explain
the (rather) sharp transition between the 
polarized and the homogeneous states observed numerically 
in Subsection~\ref{s:num-din}.
Our main result is equation \eqref{eq:tl060} below. This provides 
the hypersurface at which the transition between the polarized and the 
homogeneous state occurs in the parameter space $r$--$w$--$\ell$--$T$ for 
fixed number of particles $N$. 

The key idea is to estimate the leaking current 
$J_\textup{l}(n)$ 
emerging from a reservoir containing $n$ particles. 
Consider a time interval of length $t$ and partition it 
in smaller intervals of length $\tau=\ell\pi/(4v)$, the typical time 
a particle takes to cross the gate. 
Moreover, partition each of such time intervals in $\tau/\delta$
very small intervals of duration $\delta$  such that $v\delta\ll w$. 
As remarked in Subsection~\ref{s:efreg}, the probability that 
one particle in the urn enters the gate during a time interval of length 
$\delta$ is $p_\delta=wv\delta/(\pi A)$. 
Thus, the probability that $s$ particles enter the gate in 
the time interval of width $\tau$ is 
\begin{equation}
\label{eq:tl000}
{n\tau/\delta\choose s}p_\delta^s(1-p_\delta)^{n\tau/\delta-s}
\;.
\end{equation}
Since
$p_\delta n\tau/\delta$ is equal to the constant $nw\ell/(4 A)$, 
by the Poisson limit theorem, in the $\delta\to0$ limit, the probability 
\eqref{eq:tl000} tends to 
\begin{equation}
\label{eq:tl010}
\frac{\lambda^s}{s!}\,e^{-\lambda}
\;\;\textup{ with }\;\;
\lambda=n\frac{w\ell}{4 A}
\;.
\end{equation}
Moreover, 
the probability that at most $T$ particles enter the gate during 
the time interval of width $\tau$ is 
\begin{equation}
\label{eq:tl020}
P_\tau
=\sum_{s=0}^T\frac{\lambda^s}{s!}\,e^{-\lambda}
=\frac{\Gamma(T+1,\lambda)}{T!}
\;,
\end{equation}
where we recall the definition of the Euler incomplete $\Gamma$ 
function 
\begin{equation}
\label{eq:tl030}
\Gamma(y,x)
=
\int_x^\infty t^{y-1}e^{-t}\,\textup{d}t
\; \ y>0.
\end{equation}
Thus, typically, only in $P_\tau t/\tau$ intervals, out of the total $t/\tau$,
the particles entering the gate will not bounce back because of the threshold mechanism. 
The number of particles that will cross the gate can be 
estimated as the typical number of particles  that enter the 
gate in an interval of length $\tau$ conditioned to the fact that 
such a number is smaller than $T$ multiplied times $P_\tau t/\tau$.
Hence, the leaking current from an urn with $n$ particles is given by
\begin{equation}
\label{eq:tl040}
J_\textup{l}(n)
=
\frac{1}{t}
\frac{P_\tau t}{\tau}
\frac{1}{P_\tau}
\sum_{s=0}^Ts \frac{\lambda^s}{s!}\,e^{-\lambda}
=
\frac{4v}{\ell\pi} 
\sum_{s=0}^Ts \frac{\lambda^s}{s!}\,e^{-\lambda}
=
\frac{4v}{\ell\pi} 
\lambda
\sum_{k=0}^{T-1} \frac{\lambda^k}{k!}\,e^{-\lambda}
=
n\frac{wv}{\pi A} 
\frac{\Gamma(T,\lambda)}{(T-1)!}
\;,
\end{equation}
where we used the change of variables $s-1=k$. We remark that \eqref{eq:tl040} corresponds to a stationary
average. We then conclude that the transition line between the polarized and the homogeneous state is given 
by the equation:
\begin{equation}
\label{eq:tl060}
\frac{\textup{d}J_\textup{l}}{\textup{d}n}\bigg|_{\chi=0}
=
\frac{w v}{\pi A}\frac{\Gamma(T,\lambda)-\lambda^Te^{-\lambda}}{\Gamma(T)}
\bigg|_{\chi=0}
=0,
\end{equation}
where we used that, for any positive integer $T$, $\Gamma(T)=(T-1)!$.

\begin{figure}[t]
\centering
\includegraphics[width = 0.5\textwidth]{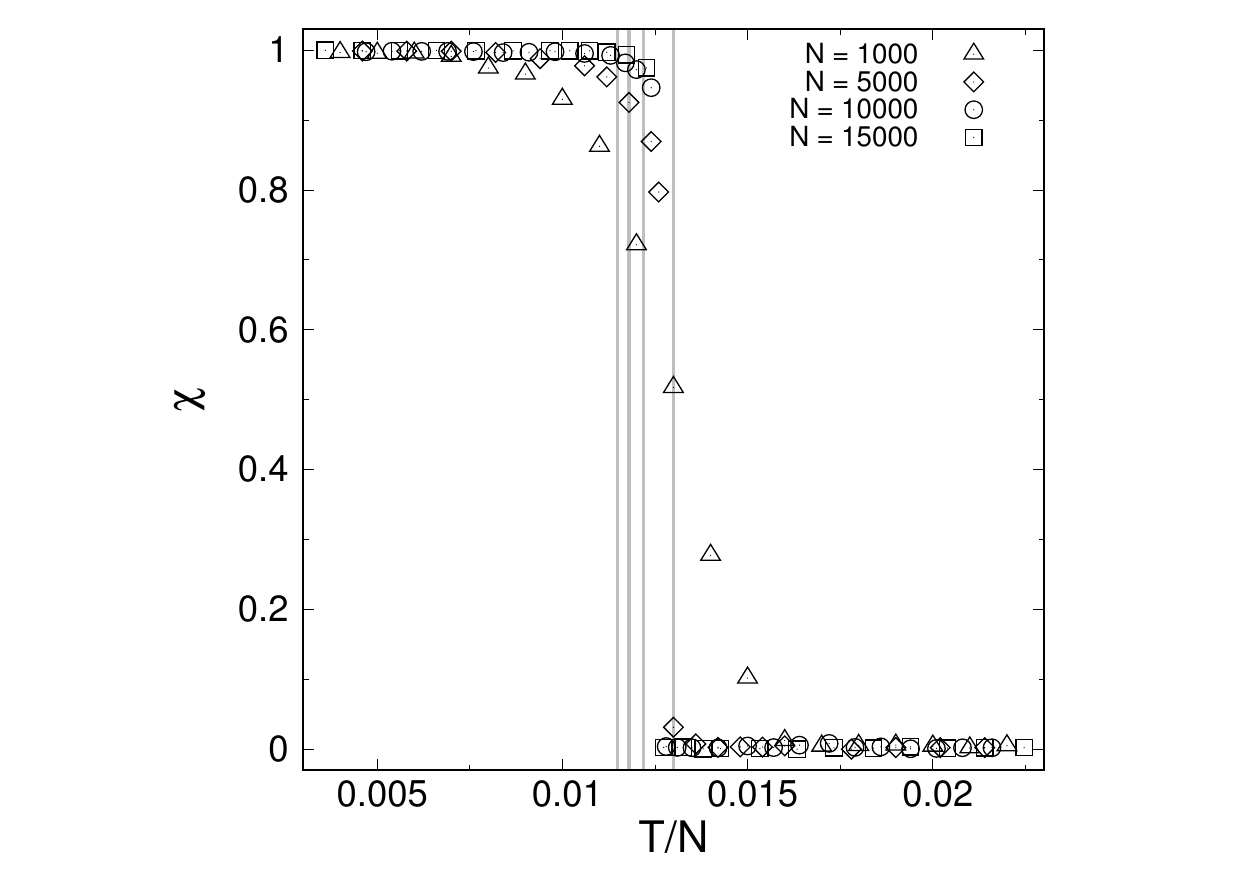} 
\caption{Mass displacement $\chi$ as a function of $T/N$, for growing $N$. 
The mass displacement for different values of $N$ 
has been measured following the procedure described in Section~\ref{s:num-din}
for $r=1$, $\ell=0.5$, $w=0.5$ and $T$ and $N$ chosen as 
described in the picture labels. 
The four grey vertical lines denote the position of the transition
estimated using \eqref{eq:tl060} for 
$N=1.5\times10^4,1.0\times10^4,5.0\times10^3,1.0\times10^3$
from the left to the right, respectively.
}
\label{ToverN}
\end{figure}

To understand the nature of the transition, let us observe Figure~\ref{ToverN}. As initially 
conjectured, it appears that this transition is determined by the growth of $T/N$, and that it 
is sharper at larger $N$. 
The transition from homogeneous 
to polarized steady states occurs when $T/N$ decreases and crosses a 
critical value that depends on the other parameters of 
the model, cf.\  equation \eqref{det000}. 
Moreover, the transition is sharper if $N$ is larger. The vertical 
lines represent the transition
value obtained from \eqref{eq:tl060} for the different values of $N$ 
considered in the picture. The larger $N$, the smaller 
the transition
value, which appears to converge to a finite 
number in the $N \to \infty$ limit.

\vfill\eject
\subsection{Estimate of the mass displacement}
\label{s:cmdis}
\par\noindent
In this section we derive an approximate closed implicit expression for the mass displacement in the polarized state.
We proceed as follows: We consider the system in the polarized state and 
suppose that the right urn is the highly populated one. Consistently with our intuition, this is
realized if the right urn is initially sufficiently more populated than the left urn, {\em e.g.}\ 
because a fluctuation has produced this situation. Therefore, in the following, we only consider 
situations in $N_R(0) > N_L(0)$, which implies $N_R(t) > N_L(t)$ for times $t > 0$.
Then, in the stationary state, the outlet
current from the left urn equals the leak current from the right reservoir, 
which, recalling
the definition of outlet current given in Section~\ref{s:efreg}
and that of leaking current given in \eqref{eq:tl040}, amounts to
\begin{equation}
\label{eq:md000}
\frac{N_\textup{L}wv}{\pi A}
=
N_\textup{R}\frac{wv}{\pi A} 
\frac{\Gamma(T,\lambda_\textup{R})}{(T-1)!}
\;,
\end{equation}
where $\lambda_\textup{R}=N_\textup{R}w\ell/(4 A)$. This leads to
\begin{equation}
\label{eq:md010}
\chi
=
\frac{1-\Gamma(T,\lambda_\textup{R})/(T-1)!}
     {1+\Gamma(T,\lambda_\textup{R})/(T-1)!},
\end{equation}
where $\lambda_\textup{R}$ depends on $\chi$ via $N_R$.  

\section{The stochastic model}
\label{s:sto}
\par\noindent
In the stochastic Ehrenfest urn model, $N$ balls are initially placed in
two urns. At each discrete time one of the two urns is chosen 
with a probability proportional to the number of particles it contains and 
one of its particles is moved to the other urn. If we denote by 
$n$ the number of particles in one of the two urns, say the first, 
the model is a discrete time Markov chain on the state space 
$\{0,1,\dots,N\}$ with transition matrix $p_{n,n+1}=(N-n)/N$ and 
$p_{n,n-1}=n/N$.

To mimic the billiard dynamics of 
Section~\ref{s:introduzione}, we modify the Ehrenfest stochastic model
introducing a threshold $T\in\{0,1,\dots,N-1\}$ that reduces the transition rates
from one urn, when its number of particles exceeds $T$. More precisely, we pick 
$\varepsilon\in(0,1]$ and, 
for $T<N/2$, we set 
\begin{equation}
\label{sto000}
p_{n,n+1}=\frac{N-n}{N}\varepsilon
\;\;\textrm{ and }\;\;
p_{n,n-1}=\frac{n}{N} \, , \quad \mbox{for $n\le T$,}
\end{equation}
\begin{equation}
\label{sto010}
p_{n,n+1}=\frac{N-n}{N}\varepsilon
\;\;\textrm{ and }\;\;
p_{n,n-1}=\frac{n}{N}\varepsilon  \, , \quad \mbox{for $T<n<N-T$,}
\end{equation}
\begin{equation}
\label{sto020}
p_{n,n+1}=\frac{N-n}{N}
\;\;\textrm{ and }\;\;
p_{n,n-1}=\frac{n}{N}\varepsilon  \, , \quad \mbox{for $n\ge N-T$}.
\end{equation}
\vskip 5pt
On the other hand, 
for $T>N/2$, we set:
\vskip 1pt
\begin{equation}
\label{sto030}
p_{n,n+1}=\frac{N-n}{N}\varepsilon
\;\;\textrm{ and }\;\;
p_{n,n-1}=\frac{n}{N} \, , \quad \mbox{for $n\le N-T$,}
\end{equation}
\begin{equation}
\label{sto040}
p_{n,n+1}=\frac{N-n}{N}
\;\;\textrm{ and }\;\;
p_{n,n-1}=\frac{n}{N} \, , \quad \mbox{for $N-T<n<T$,}
\end{equation}
\begin{equation}
\label{sto050}
p_{n,n+1}=\frac{N-n}{N}
\;\;\textrm{ and }\;\;
p_{n,n-1}=\frac{n}{N}\varepsilon \, , \quad \mbox{for $n\ge T$.}
\end{equation}
The standard stochastic Ehrenfest urn model is recovered for $\varepsilon=1$.

\subsection{The stationary measure}
\label{s:inv}
\par\noindent
The stationary measure for the Markov chain can be computed relying on  
the reversibility condition 
\begin{equation}
\label{sto100}
\mu(n)p_{n,n+1}=\mu(n+1)p_{n+1,n}
\end{equation}
for any $n=0,\dots,N-1$.

To compute the stationary measure we have to distinguish two cases. 
For $T<N/2$ the equality \eqref{sto100} can be rewritten as follows:
\begin{align}
\label{sto110}
&\mu(n)\varepsilon\frac{N-n}{N}
=
\mu(n+1)\frac{n+1}{N},
\;\;
n\le T-1
\nonumber
\\
&\mu(T)\varepsilon\frac{N-T}{N}
=
\mu(T+1)\varepsilon\frac{T+1}{N}.
\;\;
n=T
\\
&\mu(n)\varepsilon\frac{N-n}{N}
=
\mu(n+1)\varepsilon\frac{n+1}{N},
\;\;
T<n\le N-T-1
\nonumber
\\
&\mu(n)\frac{N-n}{N}
=
\mu(n+1)\varepsilon\frac{n+1}{N},
\;\;
n\ge N-T.
\nonumber
\end{align}
Solving  the recurrence relations \eqref{sto110} by induction,  we find 
the stationary measure 
\begin{align}
\label{sto120}
&
\mu(n)={\varepsilon^n \over Z} 
\left(\newatop{N}{n}\right),
\;\;n\le T
\nonumber
\\
&
\mu(n)={\varepsilon^n \over Z} 
\left(\newatop{N}{n}\right),
\;\;T+1\le n\le N-T
\\
&
\mu(n)=\frac{\varepsilon^T}{Z \varepsilon^{n-(N-T)}}
\left(\newatop{N}{n}\right),
\;\;n\ge N-T+1,
\nonumber
\end{align}
where $Z$ is the normalization constant.
On the other hand, 
for $T>N/2$, the equality \eqref{sto100} can be rewritten as follows:
\begin{align}
\label{sto130}
&\mu(n)\varepsilon\frac{N-n}{N}
=
\mu(n+1)\frac{n+1}{N},
\;\;
n\le N-T-1
\nonumber
\\
&\mu(N-T)\varepsilon\frac{N-(N-T)}{N}
=
\mu(N-T+1)\frac{N-(N-T)+1}{N}.
\;\;
n=N-T
\\
&\mu(n)\frac{N-n}{N}
=
\mu(n+1)\frac{n+1}{N},
\;\;
N-T<n< T-1
\nonumber
\\
&\mu(n)\frac{N-n}{N}
=
\mu(n+1)\varepsilon\frac{n+1}{N},
\;\;
n\ge T-1.
\nonumber
\end{align}
Solving the recurrence relations \eqref{sto130} by induction,  we find 
the stationary measure 
\begin{align}
\label{sto140}
&
\mu(n)={\varepsilon^n \over Z} 
\left(\newatop{N}{n}\right),
\;\;n\le N-T+1
\nonumber
\\
&
\mu(n)={\varepsilon^{N-T+1} \over Z}
\left(\newatop{N}{n}\right),
\;\;N-T+2\le n\le T-1
\\
&
\mu(n)=\frac{\varepsilon^{N-T+1}}{Z \varepsilon^{n-T+1}}
\left(\newatop{N}{n}\right),
\;\;n\ge T,
\nonumber
\end{align}
where $Z$ is the normalization constant.

\subsection{The large $N$ behavior}
\label{s:lar}
\par\noindent
We are interested in the structure of  the stationary measure as $N$  is large.  
We set $T=\lfloor\lambda N\rfloor$ with $\lambda\in(0,1)$.
We define 
\begin{equation}
\label{sto400}
I(n)=-\frac{1}{N}\log\mu(n)
\end{equation}
for $n=0,1,\dots,N$ and look for 
its minima. 

We discuss in detail the case $\lambda>1/2$.
In the first region, $n\le N-T+1$, simple algebraic calculations yield
\begin{equation}
\label{sto410}
I(n+1)-I(n)
=
-\frac{1}{N}
\Big[
     \log\varepsilon+\log\frac{N-n}{n+1}
\Big]
.
\end{equation}
Since
\begin{equation}
\label{sto420}
I(n+1)-I(n)>0
\Longleftrightarrow
n>\frac{\varepsilon N-1}{1+\varepsilon}
,
\end{equation}
$I(n)$ has a local minimum at 
$n^\star=\lfloor(\varepsilon N-1)/(1+\varepsilon)\rfloor$ 
in the region $0\le n \le N-T+1$ provided
\begin{equation}
\label{sto430}
\varepsilon N>1
\;\;\textrm{ and }\;\;
\varepsilon< \varepsilon^* = \frac{1}{\lambda}-1.
\end{equation}
The first  inequality is obvious. Concerning the second, we note that
$(\varepsilon N-1)/(1+\varepsilon)<N-\lambda N +1$ is equivalent to 
$[\lambda(1+\varepsilon)-1]N<2+\varepsilon$, which can be satisfied for 
arbitrarily large $N$ only if the square bracket in the left hand side 
is negative.  

With a similar argument we find that in the second region, $N-T+2\le n\le T-1$, 
the function $I(n)$ has a local minimum at 
$n^{\star\star}=\lfloor(N-1)/2\rfloor$.
In the third region, $n\ge T$, results are like in the first region, taking $N-n$ in 
place of $n$.

Finally, we compare the values attained by the function $I(n)$ at the 
two local minima $n^\star$ and $n^{\star\star}$. Assuming $N$ to be large, 
so that, in particular, the approximations 
$n^\star\approx\varepsilon N/(1+\varepsilon)$ and 
$n^{\star\star}\approx N/2$ are valid, we get 
\begin{equation}
\label{sto440}
I(n^\star)-I(n^{\star\star})<0
\Longleftrightarrow
2\varepsilon^{1-\lambda}<1+\varepsilon.
\end{equation}

\begin{figure}[t]
\centering
\includegraphics[width = 0.3\textwidth]{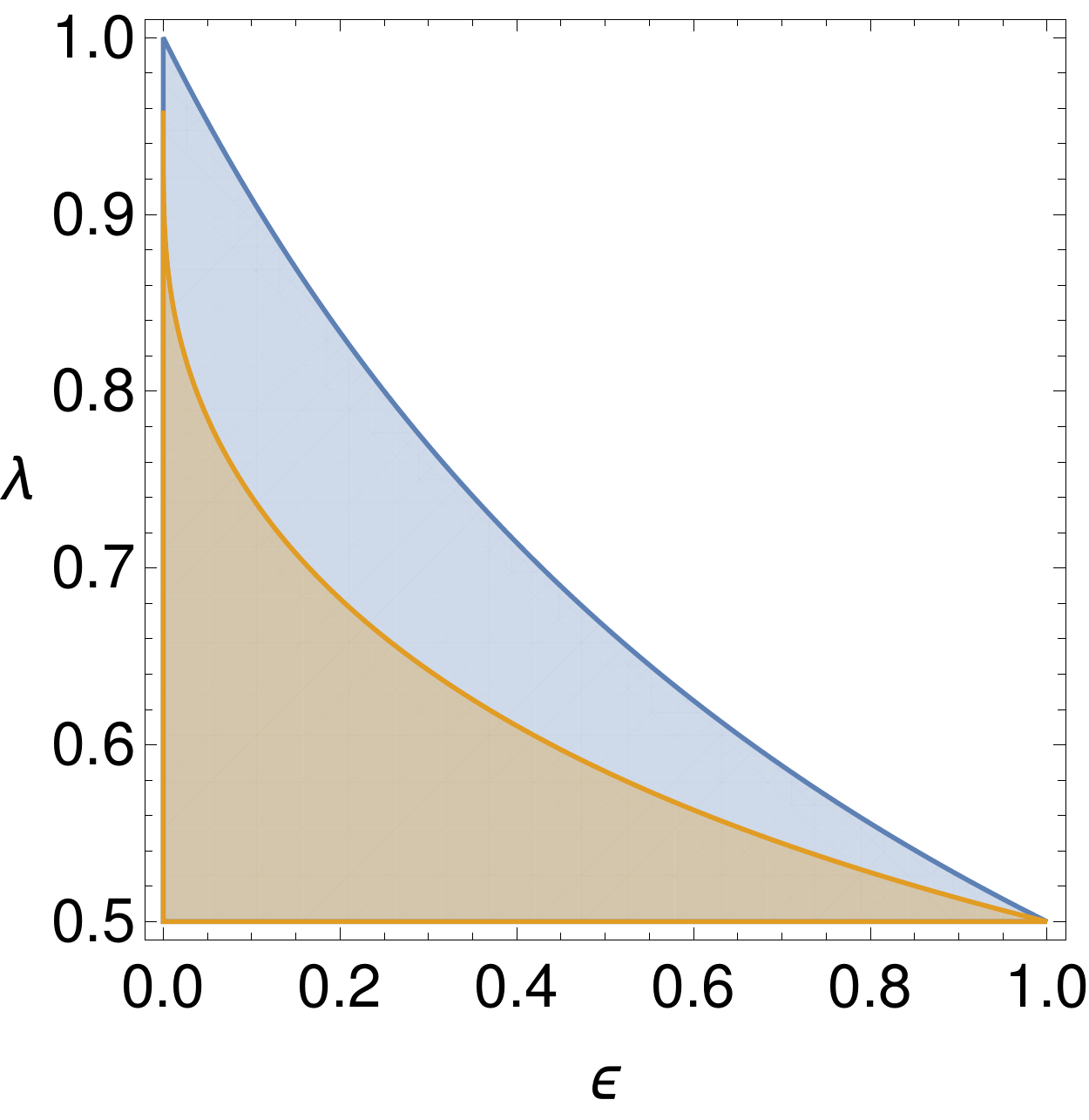} 
\hspace{2cm}
\includegraphics[width = 0.3\textwidth]{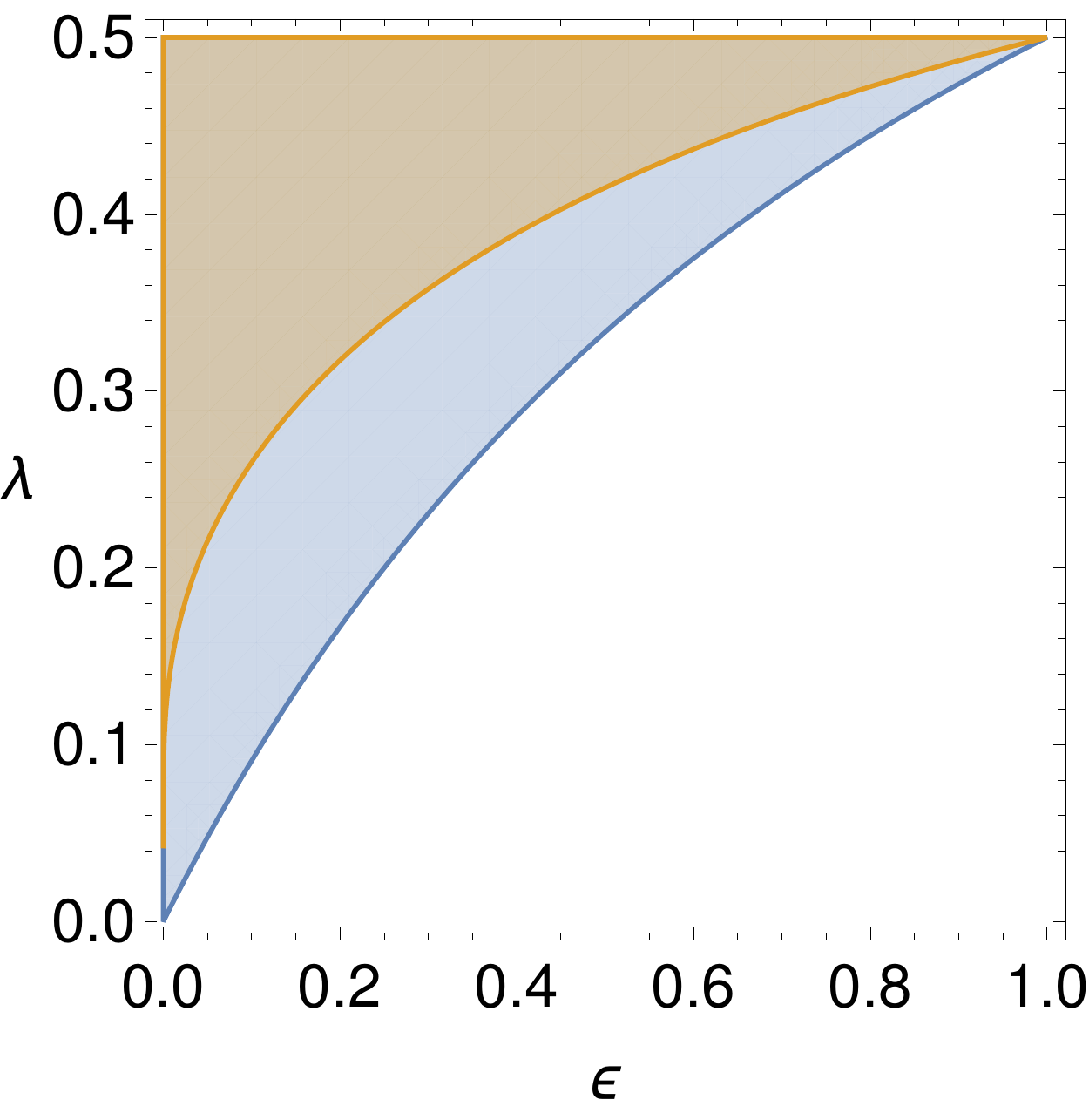} 
\caption{Left panel: Regions where conditions \eqref{sto440} 
(small bronze region) and 
the condition on the right in \eqref{sto430} (large blue region) 
are satisfied. 
Right panel:
Regions where conditions \eqref{sto460} (small bronze region) and 
the condition on the right in \eqref{sto450} (large blue region) 
are satisfied. 
}
\label{fig:fig01}
\end{figure}

As shown in the left panel of  
Figure~\ref{fig:fig01}, condition \eqref{sto440} is stronger 
than the condition on the right in equation \eqref{sto430}.
Hence we have the following options: if condition \eqref{sto440} is satisfied, 
the stationary measure concentrates, in the large $N$ limit, 
on the polarized  state, in which $n=n^\star$ or $n=N-n^\star$ 
(smaller bronze region in the left panel of Figure~\ref{fig:fig01}). If the condition on the 
right in \eqref{sto430} is satisfied, but \eqref{sto440} is violated
(part of the parameter space between the large blue and the small 
bronze region in the left panel of 
Figure~\ref{fig:fig01}), then the stationary measure 
concentrates, in the large $N$ limit, on the homogeneous 
state, in which $n=N/2$.
The states with $n=n^\star$ or $n=N-n^\star$ are sort of 
metastable states. 
For different values of the parameters the measure simply 
concentrates on the state $n=N/2$.


For $\lambda<1/2$ the discussion is similar. In this case, 
conditions \eqref{sto430} and \eqref{sto440} are respectively 
replaced by 
\begin{equation}
\label{sto450}
\varepsilon N>1
\;\;\textrm{ and }\;\;
\varepsilon<\frac{\lambda}{1-\lambda}
\end{equation}
and 
\begin{equation}
\label{sto460}
I(n^\star)-I(n^{\star\star})<0
\Longleftrightarrow
2\varepsilon^\lambda<1+\varepsilon,
\end{equation}
see also the right panel in Figure~\ref{fig:fig01}.

\subsection{Numerical simulations}
\label{s:num}
\par\noindent
In this section we report on Monte Carlo simulations 
meant to test the results obtained in Section~\ref{s:lar} and to show
the existence of a sort of metastable state, close to the 
transition line defined by conditions \eqref{sto440} and \eqref{sto460}.

\begin{figure}[t]
\centering
\includegraphics[width = 0.3\textwidth]{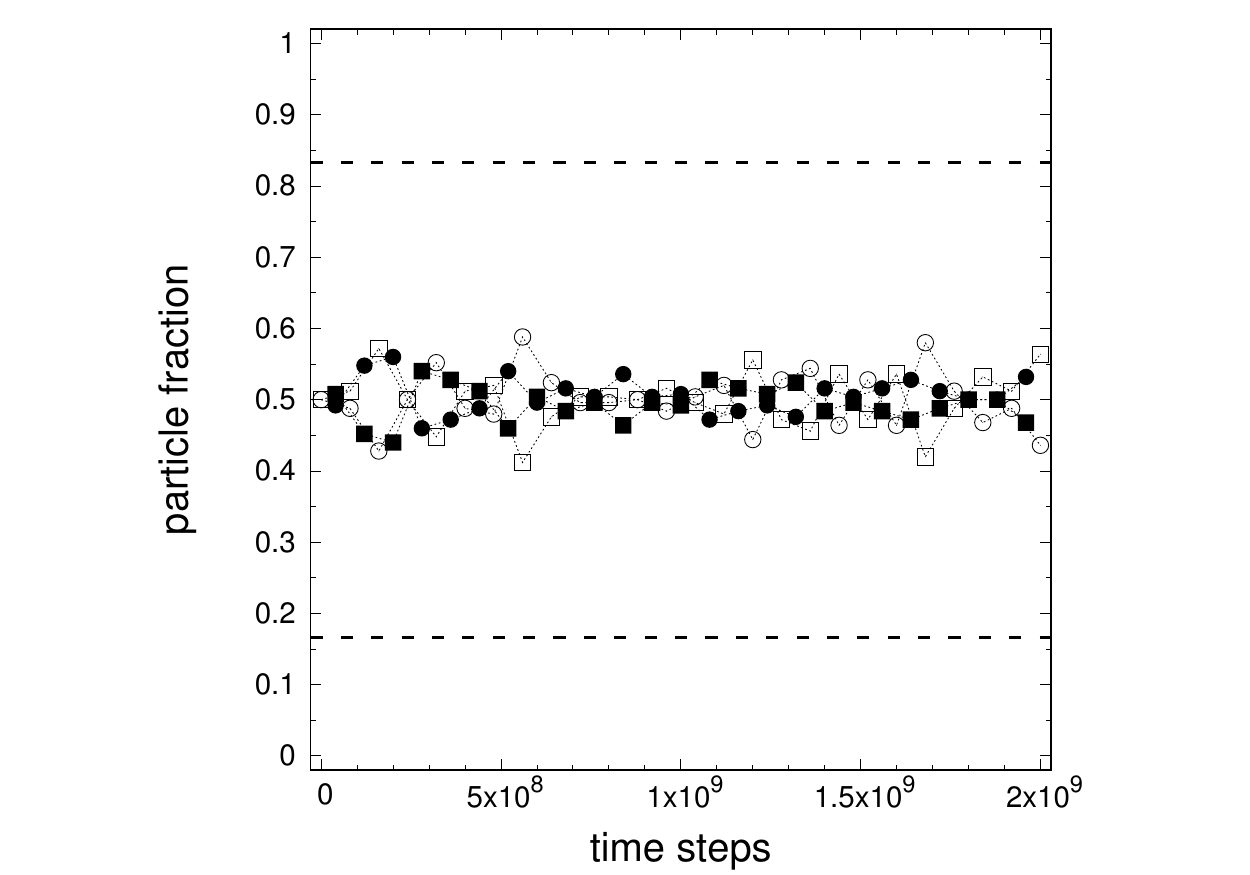} 
\hspace{0cm}
\includegraphics[width = 0.3\textwidth]{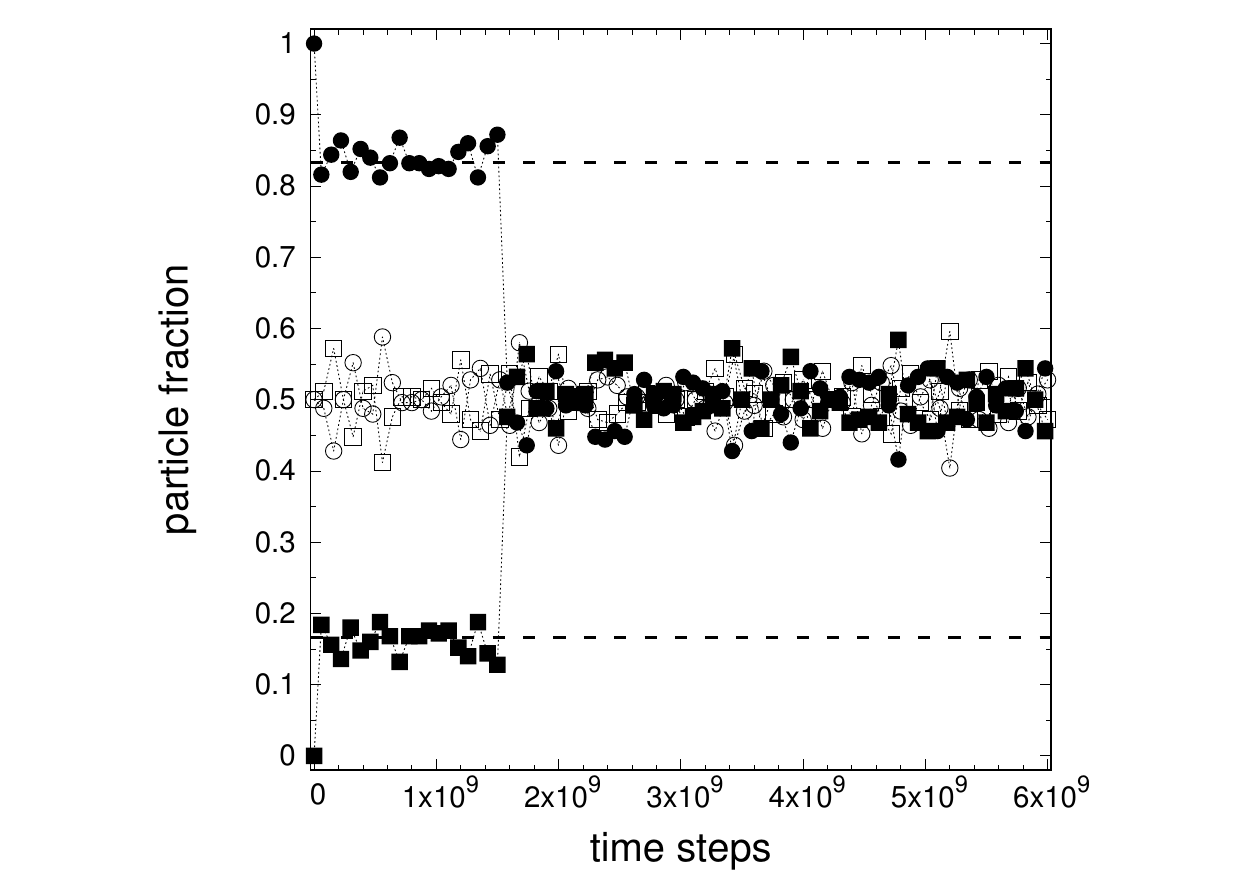} 
\hspace{0cm}
\includegraphics[width = 0.3\textwidth]{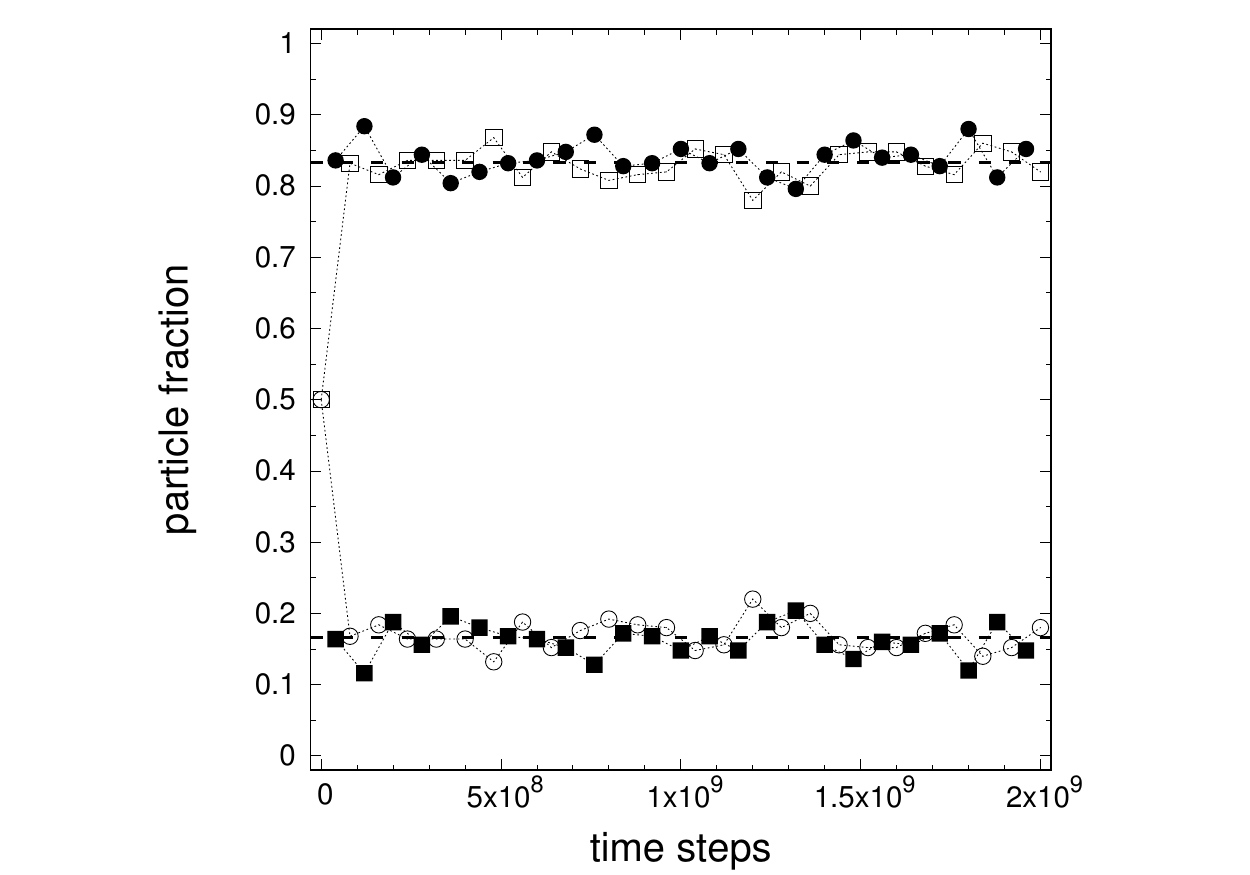} 
\\
\includegraphics[width = 0.3\textwidth]{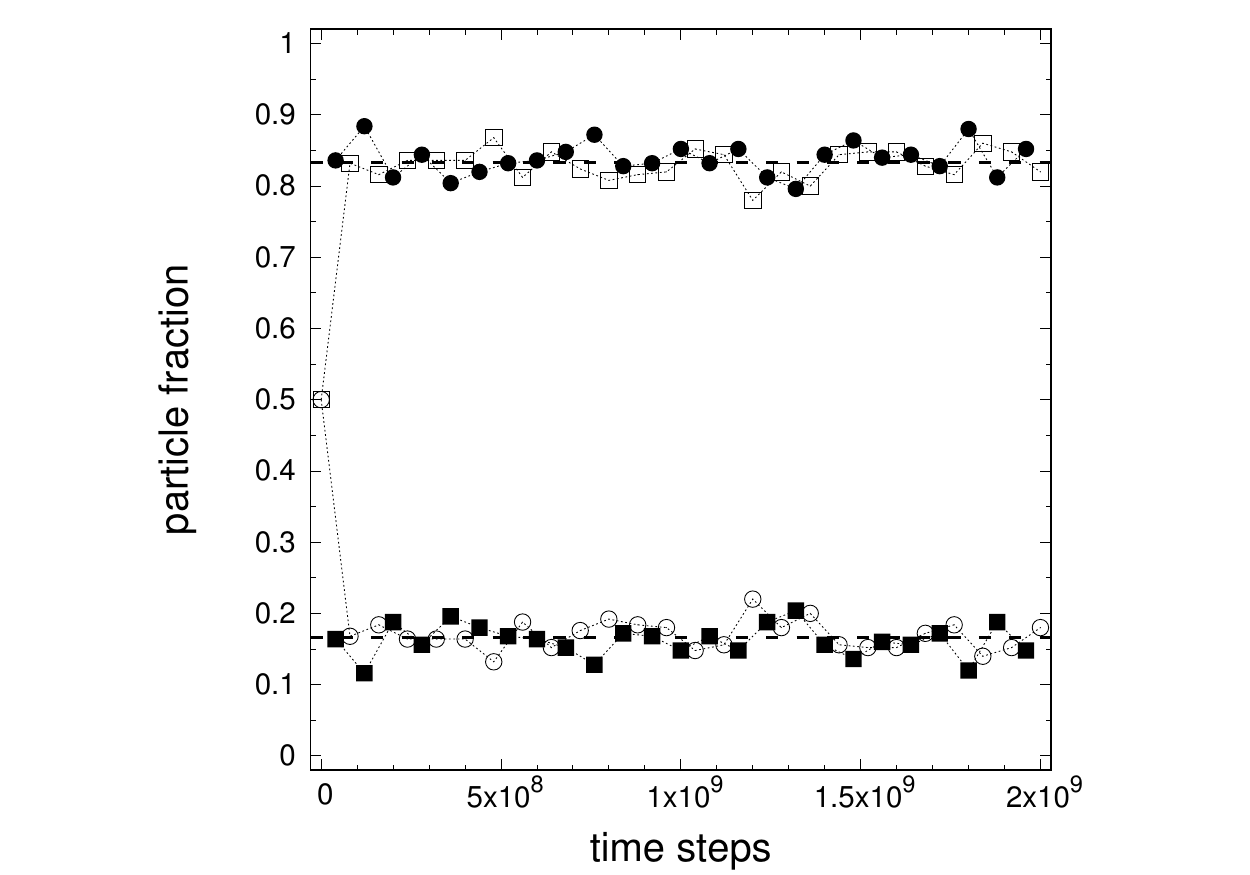} 
\hspace{0cm}
\includegraphics[width = 0.3\textwidth]{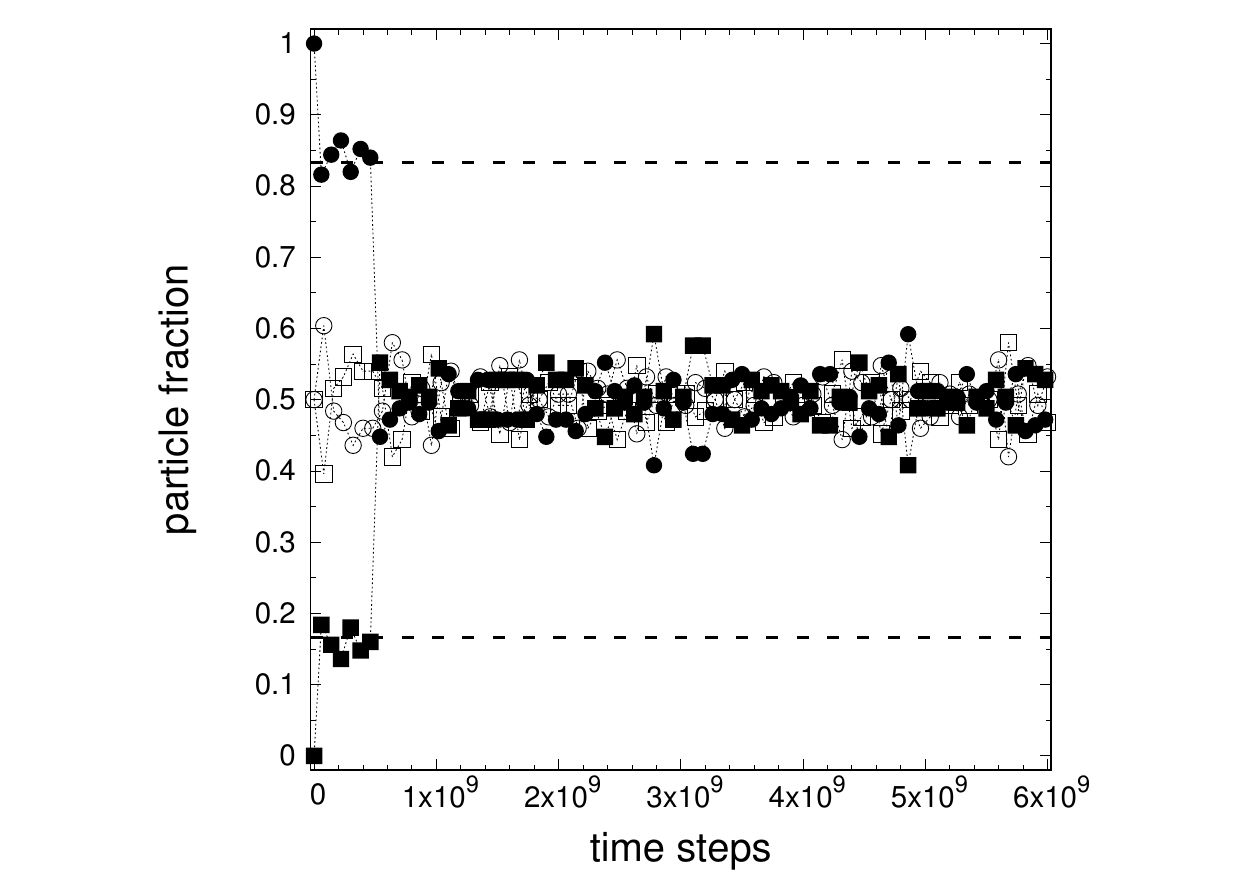} 
\hspace{0cm}
\includegraphics[width = 0.3\textwidth]{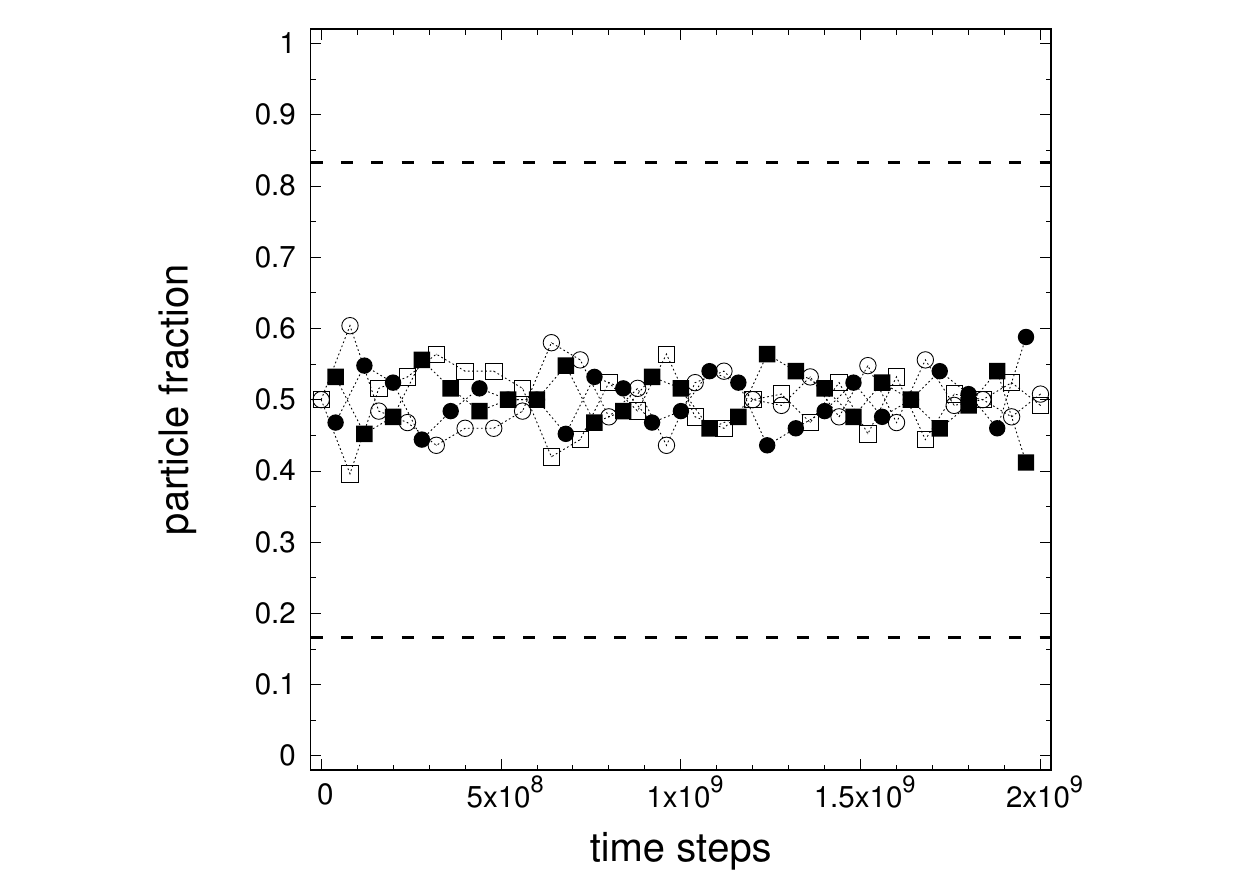} 
\caption{Number of particles in urn one (disks) and two (squares). 
Solid (open) symbols refer to the initial condition in the polarized 
(homogeneous) state $n=\varepsilon N/(1+\varepsilon)$ ($n=N/2$).
Parameters: $N=250$,
$\varepsilon=0.2$ and, in lexicographic order, 
$\lambda=0.05,0.31,0.45,0.52,0.69,0.95$.
}
\label{fig:fig02}
\end{figure}

In Figure~\ref{fig:fig02} we plot the urn content as a function of 
time for experiments with $N=250$ and $\varepsilon=0.2$. 
The initial state is either polarized (solid symbols) or 
homogeneous (open symbols). 
Looking at the picture in lexicographic order, the first 
one refers to the case $\lambda=0.005$ so that the system is represented 
by a point in the bottom (white) region in the right 
picture of Figure~\ref{fig:fig01}.
The predicted stationary state is the homogeneous one and the picture 
shows that with both initial conditions, the system quickly jumps
to such a homogeneous state and goes on fluctuating around it at any time. 
In the second picture $\lambda=0.31$ and the point representing the system 
in the right picture of Figure~\ref{fig:fig01} falls in the middle (blue) 
region, 
very close to the upper (orange) transition line. 
As confirmed by the simulation, the 
stationary state is still the homogeneous one, but if the initial 
condition is polarized, then the system spends a lot of time in such a 
state before performing, at a random time, an abrupt transition to the 
homogeneous one. 
In other words, in the middle (blue) region of Figure~\ref{fig:fig01}, very 
close to the upper (orange) transition line, the polarized state 
appears to
be a metastable state \cite{OV,CNSj}. 
In the third picture we have $\lambda=0.45$, 
and the point representing the system 
in the right panel of Figure~\ref{fig:fig01} falls in the upper (bronze) 
region. As confirmed by the simulation, the stationary state is polarized.

\begin{figure}[t]
\centering
\includegraphics[width = 0.3\textwidth]{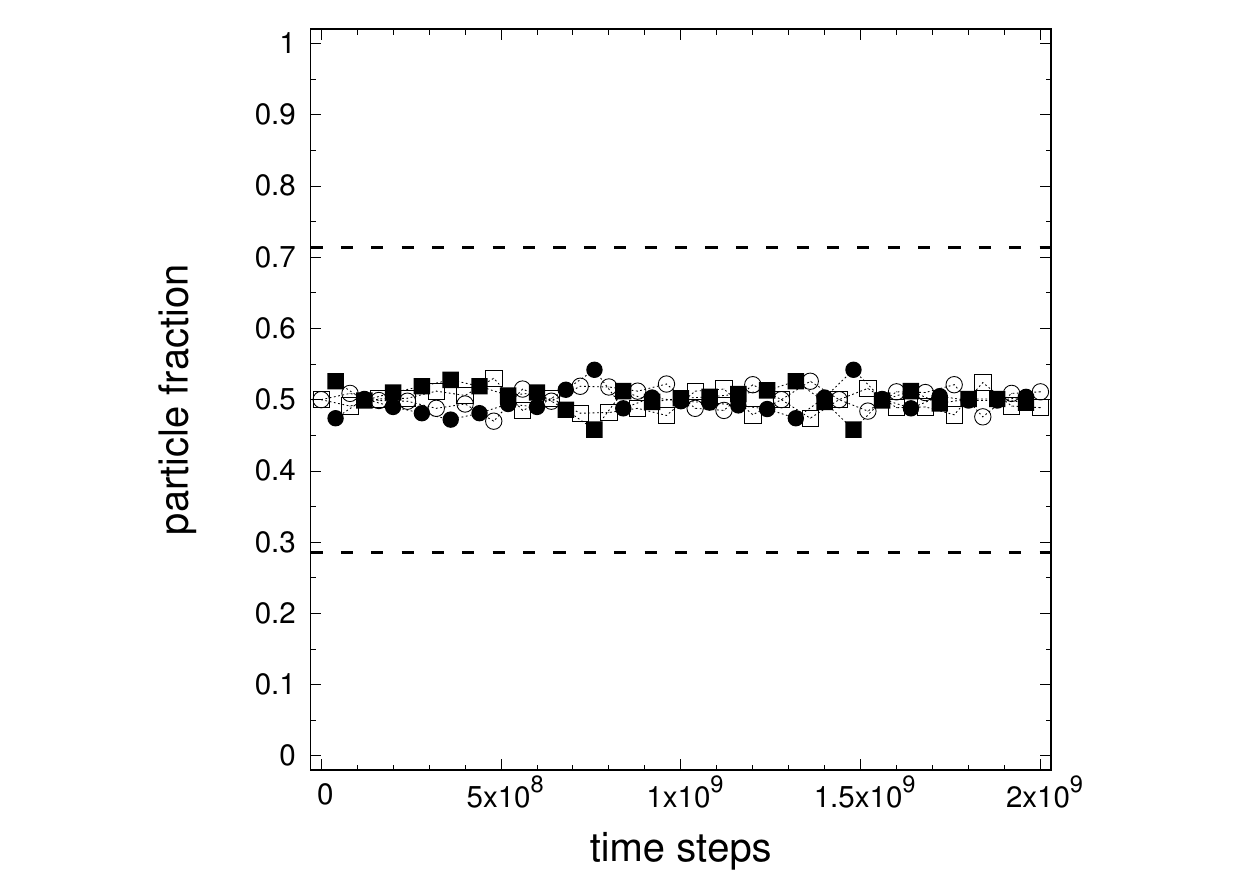} 
\hspace{0cm}
\includegraphics[width = 0.3\textwidth]{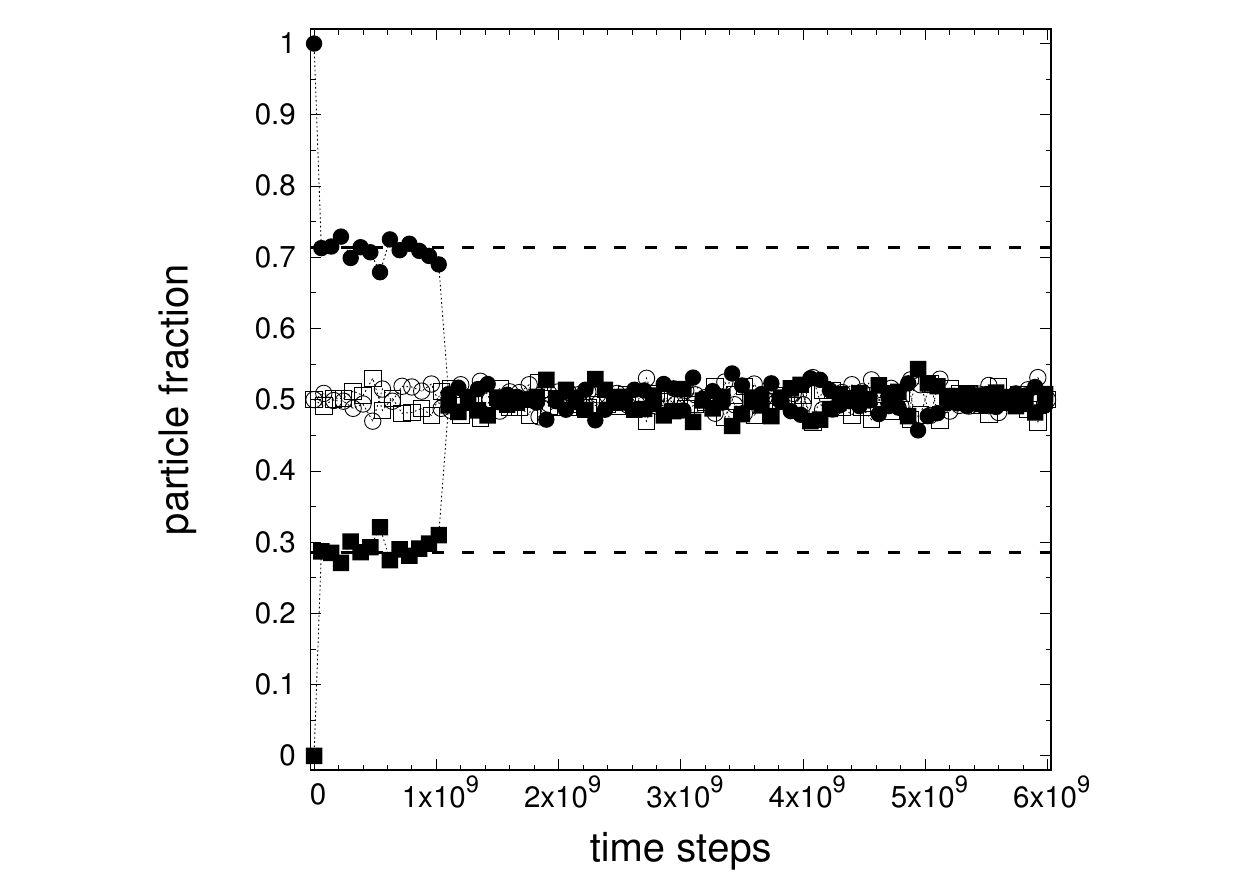} 
\hspace{0cm}
\includegraphics[width = 0.3\textwidth]{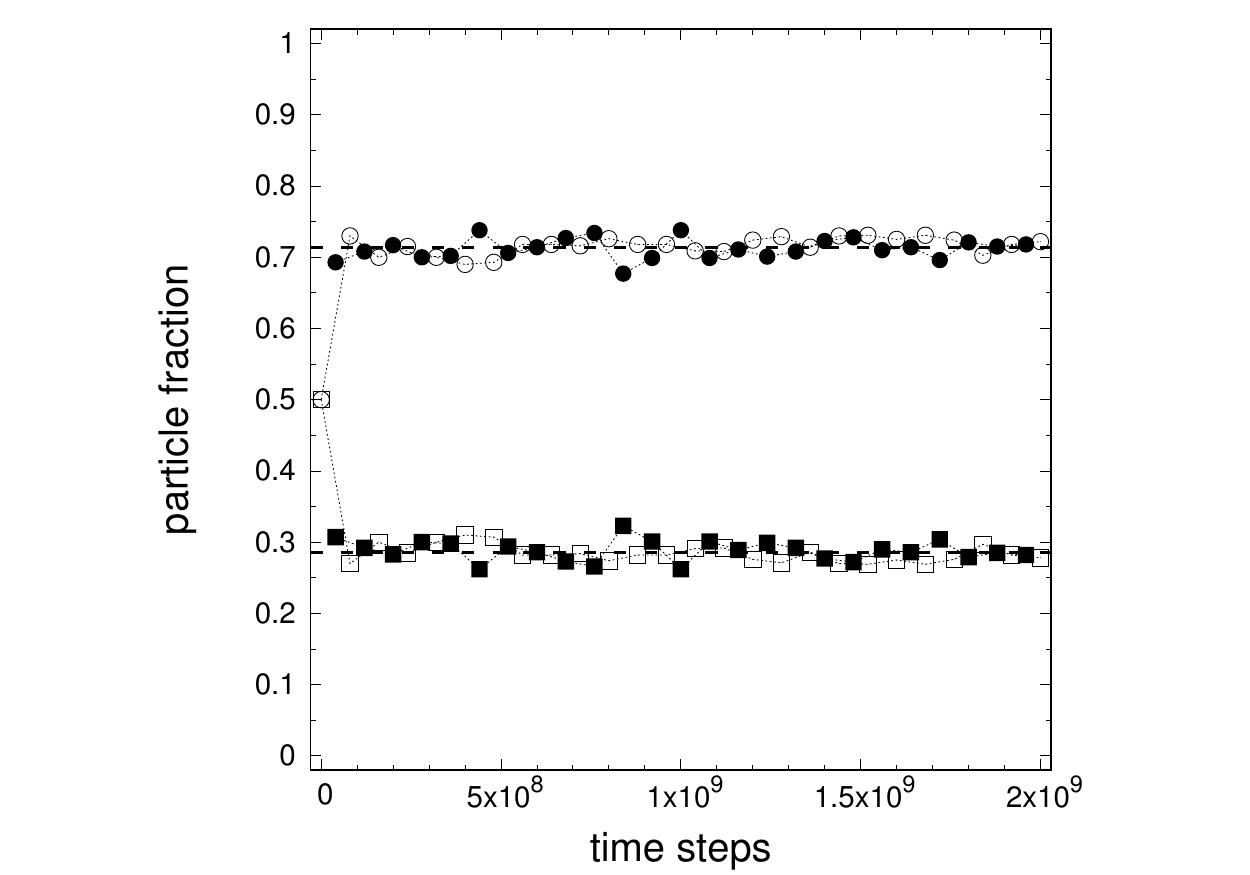} 
\\
\includegraphics[width = 0.3\textwidth]{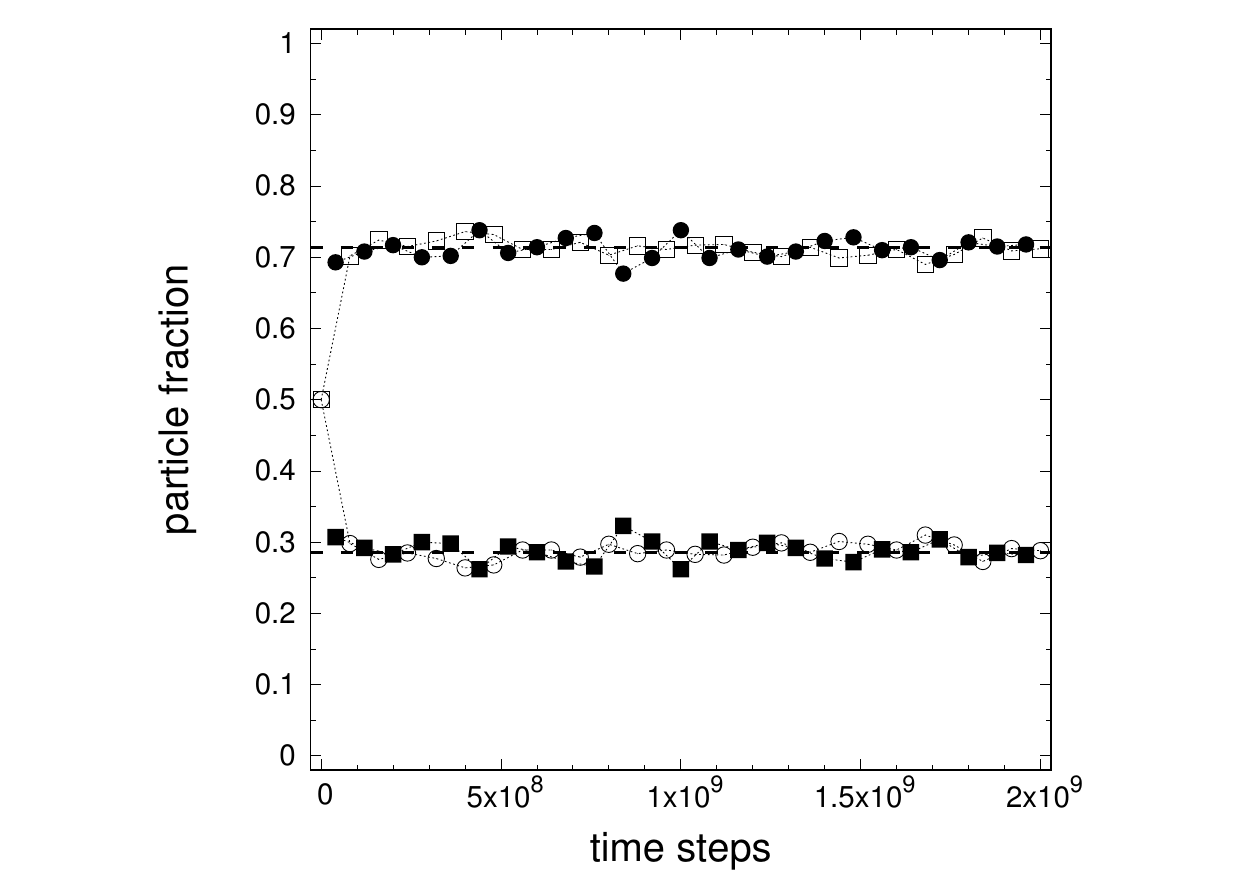} 
\hspace{0cm}
\includegraphics[width = 0.3\textwidth]{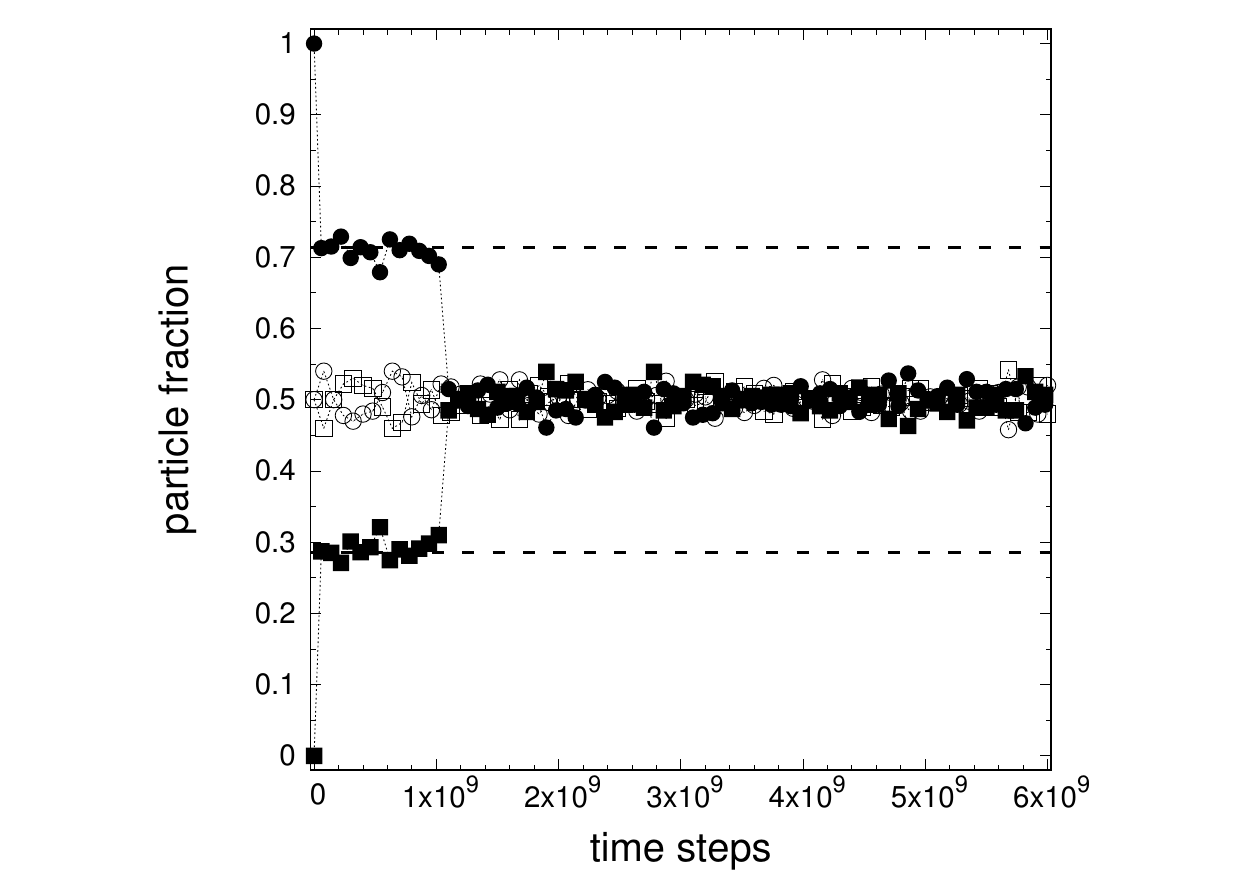} 
\hspace{0cm}
\includegraphics[width = 0.3\textwidth]{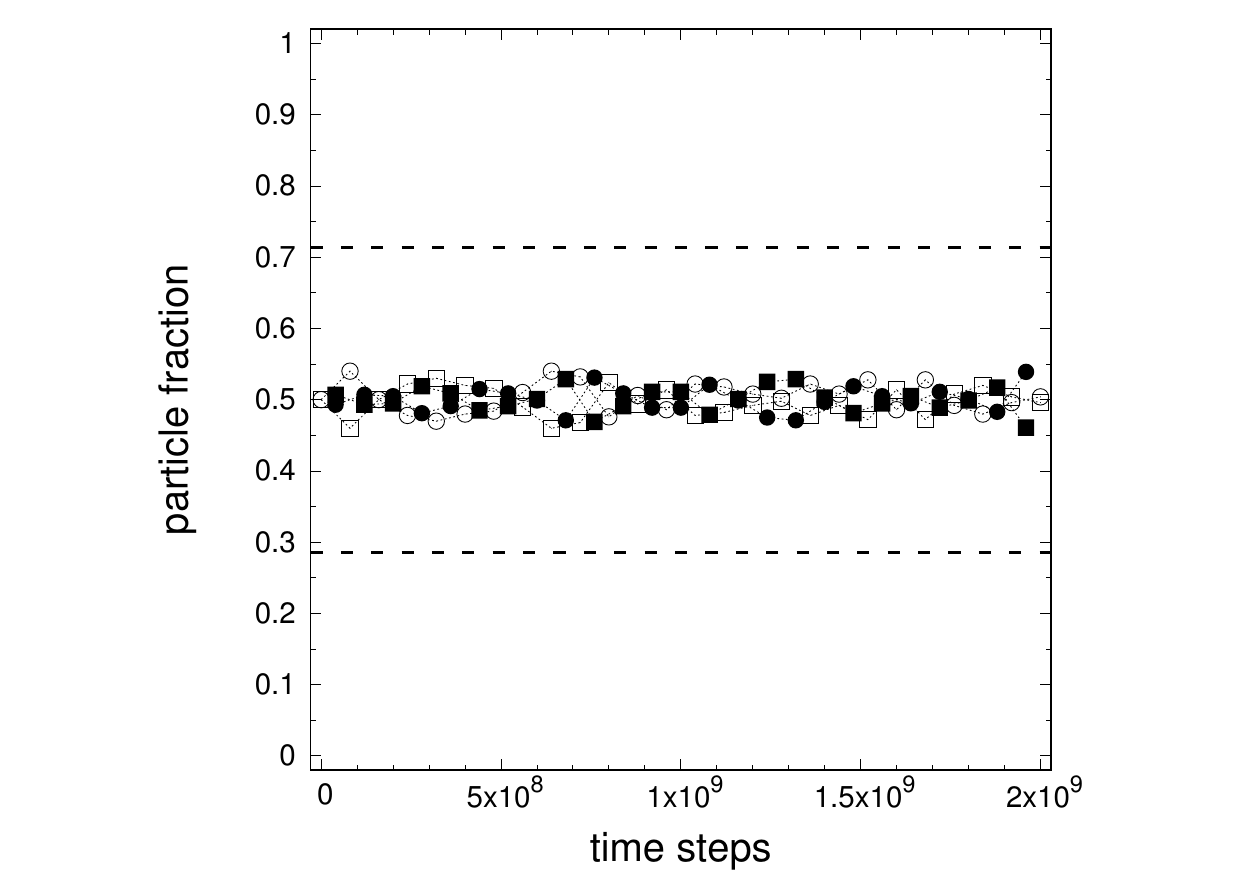} 
\caption{As in Figure~\ref{fig:fig02} with parameters
$N=1000$,
$\varepsilon=0.4$ and, in lexicographic order, 
$\lambda=0.1,0.365,0.45,0.55,0.635,0.9$.
}
\label{fig:fig03}
\end{figure}

The last three panels of Figure~\ref{fig:fig02} can be discussed similarly. 
In Figure~\ref{fig:fig03}, the analogous Monte Carlo study is repeated 
for a larger value of the parameter $\varepsilon$, namely, $\varepsilon=0.4$ and with $N=10^3$. 
Similar results are found. It is worth noticing  that the fluctuations of the numbers of particles 
around the theoretical values, indicated by the dashed lines, are smaller than those observed in 
Figure \ref{fig:fig02}. In fact, larger numbers of particles result in smaller relative fluctuations.

\section{Discussion}
\label{s:dis}
\par\noindent
In this paper, we have investigated a deterministic, TRI and phase space volume preserving particle system, 
which exhibits a non--trivial non-equilibrium phase transition, in the limit of large numbers of particles 
$N$, {\em i.e.}\ when $N$ is sufficiently large that fluctuations are negligible and 
recurrence times are unphysically long. In practice, we realized that this effect appears even at moderately large $N$, 
such as $N=O(10^4)$.
We evidenced that the transition from a homogeneous to a polarized steady state can be induced by varying the geometrical parameters of the 
billiard table (e.g. 
the radius of the urns, the width or the length of the gates). 
Moreover, the phase transition can also be determined 
by the decrease of the ratio of 
threshold and number of particles $T/N$. Remarkably, both the homogeneous and the polarized states amount to non-equilibrium steady states. 
As a matter of fact, even the homogeneous state, which means $N/2$ 
particles in the left half and $N/2$ in the right half of $\Lambda$, does not correspond to a uniform 
number density in $\Lambda$, because the density in the gates is not larger than $4T/\ell w$, which 
does not grow with $N$, while the number density in one urn is not smaller than $(N-8T)/2\pi r^2$,
which grows linearly with $N$. 
At relatively small $N$, the transition occurs gradually, over a range of $T/N$ values which shrinks
with growing $N$, eventually giving rise to a kind of first order transition. Also, larger $N$ implies 
a smaller $T/N$ value at the transition, and such a value seems to converge to a finite number in 
the $N \to \infty$ limit. Our deterministic model introduces a new kind of dynamics leading to 
non-equilibrium phase transitions apparently with no dissipation, but thanks to the action of a 
kind of Maxwell demon. Clearly, the physical implementation of the demon would imply dissipation
in the environment.

To the best of our knowledge, this is the first example of such a kind.
In the first place, most reports on non-equilibrium phase transitions concern stochastic processes and not
deterministic dynamics. Secondly, the TRI models of non-equilibrium molecular dynamics, experience phase transitions especially at low numbers of particles, when dissipation can dominate the time evolution, 
and drastic reductions of phase space volumes, not expected at large $N$ (cf. e.g. \cite{BonCohEv}), are easily 
realized.

The ergodicity of the model without the threshold mechanism suggests a stochastic counterpart. We have
thus introduced a new exactly solvable model, which is a modified version of the Ehrenfest two urns model, 
in order to match our billiard dynamics.
We have shown that the numerical results for the billiard model perfectly agree with the analytic results
of the stochastic model, once parameters are properly matched.

\end{document}